\documentclass[11pt]{article} \hoffset=-15mm \voffset=-10mm
\usepackage{graphicx,amsmath,amssymb,epsf} \usepackage{latexsym,bm,
  slashed} \usepackage{xcolor} \definecolor{dark}{rgb}{0.10,0.2,0.3}
\definecolor{magenta}{rgb}{0.7,0.1,0.3}
\definecolor{purpure}{rgb}{0.5,0.15,0.3}
\usepackage[font=small,format=plain,labelfont=bf,up,textfont=it,up]{caption}
\usepackage{hyperref, cite} 
\hypersetup{colorlinks, citecolor=blue,
  filecolor=blue, linkcolor=magenta,
  urlcolor=purpure,hyperfootnotes=true,pdftex}

 \title{\bf  \Large 
Lipatov's QCD high energy effective action: past and future      
   } \author{ Martin Hentschinski \\ \\
 Departamento de Actuaria, F{\' \i}sica y Matem{\' a}ticas,\\ \small Universidad de las Am{\' e}ricas Puebla, \\\small Santa Catarina M{\'a}rtir, 72820 Puebla, M{\'e}xico.
}

\begin{document}

\maketitle
\begin{abstract}
 In this contribution we briefly review some aspects of Lipatov’s gauge invariant QCD high energy effective action. The high energy effective provides a field theory framework to systematically factorize QCD scattering amplitudes and related theories in the limit of high center of mass energies. After a short review of the underlying concepts, we address the question how the high energy effective action can be used for actual calculations. As an explicit example we review the derivation of the gluon Regge trajectory up to 2 loop. We then review two topics of current interest: automatic amplitude generation of high energy effective action amplitudes and a discussion of the relation between the so-called Color-Glass-Condensate framework and the high energy effective action. 
 
\end{abstract}

\section{Introduction}
\label{sec:introduction}

Effective field theories are highly successful tools to study
dedicated kinematic limits of quantum field theories. This is in
particular true for the study of Quantum Chromodynamics (QCD), the
fundamental theory of strong interactions.  In 1995, Lev Lipatov
formulated a gauge invariant high energy effective action
\cite{Lipatov:1995pn,Lipatov:1996ts}, which provides the effective field theory description
of QCD in the high energy or Regge limit.  This effective action is an
effective field theory which describes the interaction between
conventional QCD degrees of freedom, {\it i.e.}  quarks, gluons and
ghost, and a newly introduced degree of freedom, the reggeized gluon.
The formulation of the high energy effective action by Lev Lipatov is
not the only effective field theory description of the QCD high energy
limit: most notably one should mention first attempts by Verlinde and
Verlinde \cite{Verlinde:1993te}, see also \cite{Arefeva:1993hi} as
well as an earlier version of an high energy effective action by Lev
Lipatov \cite{Lipatov:1991nf} and collaborators
\cite{Kirschner:1994gd,Kirschner:1994xi}. Another important framework, which has been
developed around the same time as Lipatov's high energy effective
action, is provided by the Color Glass Condensate effective theory and
its ensuing calculational framework. We will comment on  recent
results concerning the relation of both frameworks further down in the
text. \\

The benefits of such a effective field theory framework are manifold:
first and most important they provided a field theory
definition of the underlying degrees of the freedom in the $t$-channel
of high energy factorized QCD amplitudes: the reggeized gluon. While the
bare reggeized gluon fields is introduced as a new auxiliary degree in
the formulation of the high energy effective action, reggeization of
the gluon, {\it i.e.} non-zero Regge trajectory arises within the high
energy effective action as a direct consequence of the calculation of
perturbative higher order corrections to the propagator of this new
``auxiliary'' degree of freedom.  Introducing this degree of freedom
is therefore crucial for the formulation of the high energy effective
action. By definition this new degree of freedom is invariant under
local gauge transformation which in turn essential to achieve gauge
invariant factorization of scattering amplitudes in the high energy
limit.
\\

The outline of this short review is as follows: in section
Sec.~\ref{sec:high-energy-effect} we briefly review the formulation of
the high energy effective action.  In
Sec.~\ref{sec:high-energy-effect-1} we review essential ingredients
for the use of the high energy effective action in actual
calculations. In Sec.~\ref{sec:gluontrajectory-effectiveaction} we
give a short overview of the determination of the 2-loop gluon
trajectory. Sec.~\ref{sec:numerical-methods} provides a short
introduction on amplitude generation for the high energy effective
action, while in Sec.~\ref{sec:high-energy-effect-4} we address the
high energy effective action in the presence of a dense reggeized
gluon field and its relation to the framework of the Color Glass
Condensate effective theory. In Sec.~\ref{sec:conclusions} we
summarize our results and most importantly highlight those results
which we could not cover in this short review.  In particular we would
like to note that the topics covered in this review are highly biased
by the own work of the author. This is mainly due to the limitations
of the author and implies by no means that work not mentioned would be
of less relevance.

\section{The high energy effective action: derivation}
\label{sec:high-energy-effect}

\begin{figure}[t]
  \centering
  \includegraphics[width=.35\textwidth]{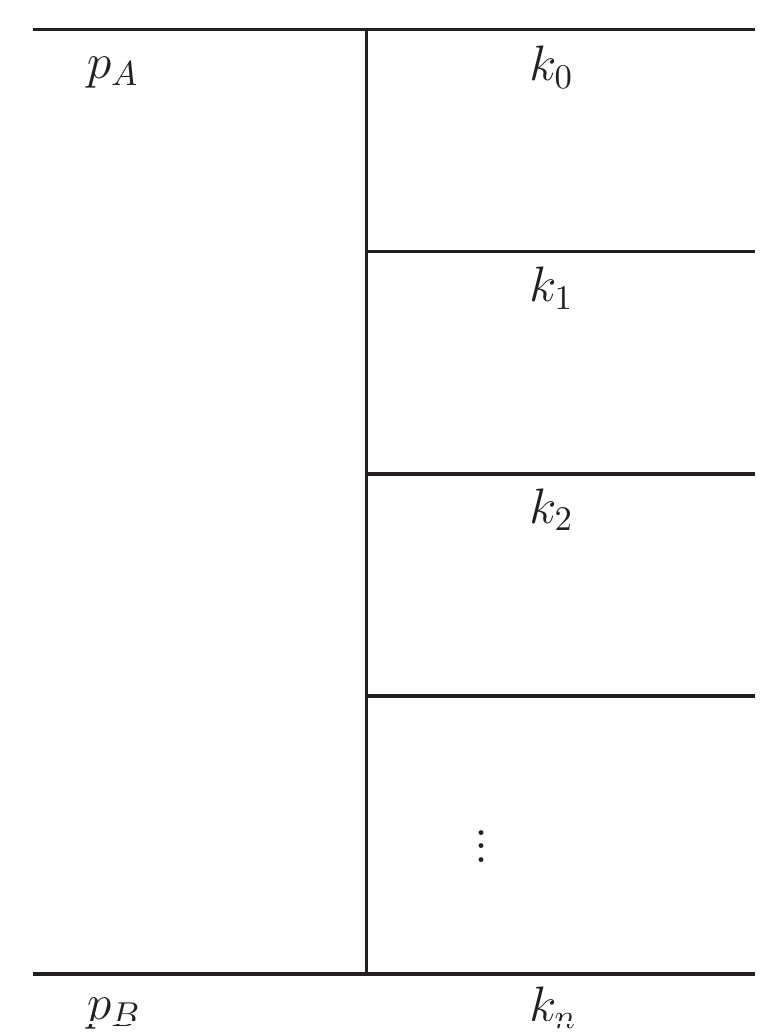}
  \caption{The Multi-Regge-Kinematics}
  \label{fig:mrk}
\end{figure}
The high energy effective action aims at a description of scattering
amplitudes which exhibit regions where plus and minus momenta are
strongly ordered. It is instructive to consider the example of an
explicit scattering process $p_A + p_B \to $ X in the high energy
limit (`any mass scale')$^2$/$s \to 0$ with $s = (p_A + p_B)^2$ the
center of mass energy squared. In this limit, the momenta of
scattering particles $p_{A,B}$ are close to the opposite sides of the
light cone. It is therefore natural to define dimensionless light-cone
vectors associated with these momenta. If $p_{A,B}$ are already
light-like, these are simply obtained through the following re-scaling
\begin{align}
  \label{eq:light-cone}
 n_\pm & = \frac{2}{\sqrt{s}} p_{B,A}, & n_+ \cdot n_- & = 2.
\end{align}
Plus- and minus components of momenta (and general four vectors) are
 defined through
\begin{align}
  \label{eq:lc_comp}
  k^\pm & = n^\pm \cdot k = n_\mp \cdot k = k_\mp.
\end{align}
This implies the following Sudakov decomposition of a four vector,
\begin{align}
  \label{eq:Sudakov}
  k & =  k^+ \frac{n^-}{2} + k^- \frac{n^+}{2} + {\bf k}  =  k_- \frac{n_+}{2} + k_+ \frac{n_-}{2} + {\bf k},
\end{align}
where ${\bf k}$ denotes the momentum perpendicular to the 2 light-cone
directions with ${\bf k}^2 > 0$. For derivatives the following
relations hold:
\begin{align}
  \label{eq:1}
  \partial_\pm x^\pm & = 2 &  \partial_\mp x^\pm & = 0. 
\end{align}
The kinematic limit of interest is then defined as follows: a genuine
scattering process is organized into several sub-sectors of particles
which are close, {\it i.e.} local, in rapidity
$\eta = \frac{1}{2} \ln \frac{k^+}{k^-}$. These sectors are strongly
ordered with respect to each other in both plus and minus momenta (and
therefore in rapidity). The situation is most easily illustrated using
the tree diagram depicted in Fig.~\ref{fig:mrk}: the momenta of the
each of the sectors $k_i$ with $i = 0, \ldots, n$ obey the conditions
$k_0^+ \gg k_1^+ \gg \ldots k_n^+$ and
$k_0^- \ll k_1^- \ll \ldots k_n^-$. As a side remark we note that for
real particle production, this condition corresponds to the well known
Multi-Regge-Kinematics {\it i.e.} strong ordering in rapidity, while
all transverse momenta are of the same order of magnitude.

\subsection{The high energy effective action for gluons}
\label{sec:high-energy-effect-2}

Following closely \cite{Lipatov:1995pn,Lipatov:1996ts}, the essential idea is to decompose
the complete gluonic field $V_\mu$ field into a sum of fields
$v^{(s)}_\mu$ and $A_\mu^{(s)}$, where the upper index `$s$' denotes a
certain sector $s$ which is local in rapidity, {\it i.e.} a sector
which has no regions with strong ordering in plus and minus
momenta. The index `$s$' indicates that the fields $v^{(s)}_\mu$ is
restricted to such a sector and for each such sector, there exists an
individual field $v^{(s)}_\mu$.  The fields $A^{s}_\mu$ differ from
these fields: while the index `$s$' indicates that these fields also
couple to a certain rapidity sector, these fields never occur
\emph{inside} a certain sector. They rather transmit the interaction
between different sectors. They are ``non-local in rapidity''.  They
generally appear in the $t$-channels of scattering processes and do
not appear as asymptotic states.  Since the fields $A^{s}_\mu$ connect
to sectors which are strongly ordered in plus and minus momenta with
respect to typical momenta in the sector $s$, their coordinate
dependence is strongly boosted with respect to the regarding
light-cone direction. As a consequence, only certain components of the
the fields $A^{s}_\mu$ are of relevance. In particular,
\begin{align}
  \label{eq:boost1}
  A^{\mu (s)}(x) & = A_+^{(s)} (x) \frac{n_-^\mu}{2} + A_-^{(s)} (x) \frac{n_+^\mu}{2},
\end{align}
while the  dependence of the fields on light-cone
coordinates is frozen: 
\begin{align}
  \label{eq:3}
  \partial_+ A^{(s)}_- (x)& = 0 = \partial_- A^{(s)}_+.
\end{align}
In the momentum representations, to which we will turn soon, this
encodes strong ordering of different sectors (which the $A_\pm$ fields
connect) with respect to their plus and minus momenta. As we will see
below, the field $A_\mu$  directly yields the so-called reggeized
gluon, which has been identified by Lipatov and collaborators as the
fundamental degree of freedom in the $t$-channel of high energy
factorized scattering amplitudes. Although the fields $A_\mu$ are
initial not reggeized, {\it i.e.} they have no non-zero Regge
trajectory, reggeization of this fields will result from the
resummation of radiative corrections. Following \cite{Lipatov:1995pn,Lipatov:1996ts}, we
therefore will generally refer to the fields $A_\mu$ as ``reggeized
gluon'' fields.
\\

The essential goal is now to construct the effective action which
describes the local  
interaction, {\it i.e.} inside the regarding sectors, between  $v^{(s)}$  and $A_\pm^{(s)}$ fields and
$v^{(s)}$ fields themselves.  The interactions among the   $v_\mu^{(s)}$
fields is described by the conventional QCD Lagrangian, with the implicit
constraint that the resulting QCD dynamics must be local  in rapidity.  In
addition to the QCD Lagrangian, one further  requires a term which describes the interaction among
$v_\mu^{(s)}$ and $A_\mu^{(s)}$ fields.  The resulting Lagrangian of
the effective theory is therefore obtained as the sum of the
conventional QCD Lagrangian and a so-called \emph{induced} Lagrangian $ \mathcal{L}_{\text{ind.}}$, which 
couples the gluon field  to the 
$A_\mu$ field:
\begin{align}
  \label{eq:2}
\mathcal{L}_{\text{eff}} & =   \mathcal{L}_{\text{QCD}}[v^{(s)}_\mu, \psi^{(s)}, \bar{\psi}^{(s)}]
  + \mathcal{L}_{\text{ind.}}[v^{(s)}_\mu, A^{(s)}_+,  A^{(s)}_+].
\end{align}
Here the QCD Lagrangian is as usually given by
\begin{align}
  \label{eq:12}
  \mathcal{L}_{\text{QCD}}[v^{(s)}_\mu, \psi^{(s)}, \bar{\psi}^{(s)}]
  & = - \frac{1}{2} \text{tr} \left[G_{\mu\nu} G^{\mu\nu} \right] +
    \sum_{f=1}^{n_f} \bar{\psi}_f (i {D\!\!\!\!/} - m_f) \psi_f.
\end{align}
In the effective Lagrangian, vectors fields are generally understood as anti-hermitian
 matrices in
 the fundamental representation of SU$(N_c)$,  {\it i.e.} $A_\pm(x) =
 - i t^a  A^a_\pm (x)$ and $v_\mu = -it^a v_\mu (x)$ with $t^a$ the
 SU$(N_c)$ generators in the fundamental representation tr$(t^at^b) =
 \delta^{ab}/2$; the trace is therefore  in the fundamental
 representation. Furthermore
 \begin{align}
   \label{eq:13}
   D_\mu & = \partial_\mu + g v_\mu,  & G_{\mu\nu} & = \frac{1}{g}
                                                   [D_\mu, D_\nu] \, .
 \end{align}
 From now on we will drop the index `$s$' and implicitly assume that
 the fields of the effective action are limited to a certain sector.
As a  first na\"ive 
attempt  to construct the induced Lagrangian, one might start with the
term
\begin{align}
  \label{eq2:efflagrangianInd1}
  \text{tr} \left [2v^\mu \partial^2 A_\mu  \right]= \text{tr}\left[ v_+ \partial^2 A_- +
  v_- \partial^2 A_+  \right],
\end{align}
which provides the simplest coupling between the 2 fields.  The
differential operator `$\partial^2$' has been introduced to ensure that
the field $A_\mu$ cannot appear as an asymptotic state: whenever the
momentum of the $A_\mu$ is the on the mass-shell, $\partial^2 = 0$,
the field will  decouple from QCD dynamics. To obtain the correct
normalization for the propagator of the $A_\mu$-field, one further
introduces an corrective term such that the coupling between both
fields reads
\begin{align}
  \label{eq2:efflagrangianInd2}
  \text{tr} \left[ (v_+ - A_+) \partial^2 A_- + (v_- - A_-) \partial^2 A_+ \right ].
\end{align}
The above term is sufficient to construct the Lagrangian of the
effective theory, as long as we ignore questions related to gauge
invariance: the term induces a transition between the conventional
gluon field $v_\mu$ inside a certain sector and the field $A_\mu$,
which connects sectors strongly ordered in rapidity with respect to
each other. The strong ordering condition is imposed on the field
$A^\mu$ by both polarization and the kinematic constraint
Eq.~\eqref{eq:3}.  For a gauge theory, the above proposal comes with
an essential deficit: factorization in the high energy limit of a
gauge independent quantity, {\it i.e.} a scattering amplitude, must
happen in an gauge independent way. To achieve this, Lipatov proposed to define the reggeized gluon fields
$A^\mu$ to be invariant under local gauge transformations
\begin{align}
  \label{eq:7}
  \delta_L A_\pm & = [A_\pm, \chi_L] = 0 ,
\end{align}
with $\chi_L$ the parameter of local gauge transformation which decreases for $x \to \infty$. QCD fields are on the other hand as usual  subject to 
such  local gauge transformations, 
\begin{align}
  \label{eq:6}
 \delta_L v_\pm & =  [D_\pm, \chi], & \delta_L \psi & = - \chi,  & D_\mu & = \partial_\mu + g v_\mu.
\end{align}
As a consequence, the proposal Eq.~\eqref{eq2:efflagrangianInd2} for
the coupling between the $v^\mu$ and the $A^\mu$ fields violates local
gauge invariance. The gauge invariant high energy effective action
proposed by Lipatov solves this problems by adding the minimal
necessary terms to re-establish gauge invariance, leading to the
following induced Lagrangian:
\begin{align}
\label{eq:4}
\mathcal{L}_{\text{ind}} [v^\mu, A_+, A_-] & = \bigg\{ 
\text{tr}\left[\left(T_-[v(x)] - A_-(x) \right)\partial^2_\perp A_+(x)\right]
\notag \\
& \hspace{3cm}
+\text{tr}\left[\left(T_+[v(x)] - A_+(x) \right)\partial^2_\perp
  A_-(x)\right] \bigg\}.
\end{align}
The functionals $T_\pm[v] $ can be obtained from the following
operator definition
\begin{align}
\label{eq2:efflagrangian}
T_\pm[v] =
&
-\frac{1}{g}\partial_\pm  \frac{1}{1 + \frac{g}{\partial_\pm}v_\pm}
 =  v_\pm - g  v_\pm\frac{1}{\partial_\pm} v_\pm + g^2 v_\pm
\frac{1}{\partial_\pm} v_\pm\frac{1}{\partial_\pm} v_\pm - \ldots ,
\end{align}
where the integral operator $1/\partial_\pm$ is implied to act on a
unit constant matrix from the left. Boundary conditions of the
$1/\partial_\pm$ are fixed through
\begin{align}
  \label{eq:6x}
 \frac{1}{1 + \frac{g}{\partial_\pm}v_\pm} &  = 
 U[v_\pm(x)] = \mathcal{P}\exp\bigg(-\frac{g}{2} \int\limits_{-\infty}^{x^\pm}dx'^\pm  v_\pm(x')\bigg)
\notag \\ 
& \hspace{-1cm}= 
1 -\frac{g}{2} \int_{-\infty}^{x^\pm}dx'^\pm  v_\pm(x') + \frac{g^2}{4} \int\limits_{-\infty}^{x^\pm}   dx^{'\pm}   \int\limits_{-\infty}^{x^{'\pm}} dx^{''\pm}   v_\pm(x') v_\pm(x'')
 + \ldots\,.
\end{align}
The functional $T[v_\pm]$ therefore generalizes the na\"ive coupling between $A_\mu$ and $v_\mu$ fields. In particular 
\begin{align}
  \label{eq:5}
  \delta_L \left( \text{tr} \left[  T[v_\pm] \partial^2 A_\mp  \right] \right)& = 0,
\end{align}
and the  induced Lagrangian Eq.~\eqref{eq:4} is therefore invariant under local
gauge transformations.

\subsection{The high energy effective action for quarks}
\label{sec:high-energy-effect-3}

Following the same ideas, one can write down the high energy effective
action for quarks, meant to describe the high energy limit of
processes with flavor exchange \cite{Lipatov:2000se}. The degree of
freedom, $Q_\pm$, which is used to factorize amplitudes in the high
energy limit, is in this case not  bosonic, but fermionic. In
correspondence to Eq.~\eqref{eq:boost1} this field satisfies:
\begin{align}
  \label{eq:8}
  {n\!\!\!/}_\pm Q_\mp & = 0, &   \bar{Q}_\mp  {n\!\!\!/}_\pm & = 0, 
\end{align}
while the dependence on light-cone coordinates is frozen,
\begin{align}
  \label{eq:9}
  \partial_\pm Q_\mp(x) & = 0 = \partial_\pm \bar{Q}_\mp(x) = 0,
\end{align}
in accordance with Eq.~\eqref{eq:3}. The complete effective Lagrangian
is given by
\begin{align}
  \label{eq:10}
  \mathcal{L}_{\text{eff}} 
       & = \mathcal{L}_{\text{QCD}} [v_\mu, \psi, \bar{\psi}]
         + \sum_{f = 1}^{n_f} \mathcal{L}_{\text{ind.}}^{(Q_f)} [\psi, \bar{\psi}, v_\mu, Q_\pm, \bar{Q}_\pm],
\end{align}
where the induced Lagrangian  $ \mathcal{L}_{\text{ind.}}^{(Q)}$ describes again the coupling of the $Q_\pm$ and $\bar{Q}_\pm$ fields to QCD fields:
\begin{align}
  \label{eq:11}
   \mathcal{L}_{\text{ind.}}^{(Q_f)}  [\psi, \bar{\psi}, v_\mu, Q_\pm, \bar{Q}_\pm]
& =
 \bar{Q}_{-}^f(x)\left(i {\partial \!\!\!/} - m_f \right) \left[Q^f_+(x) - U^\dagger[v_+(x)] \psi_f(x) \right] 
\notag \\
&  \hspace{-2.5cm} 
 +
 \left[\bar{Q}^f_+(x) - \bar{\psi}(x) U[v_+(x)] \right] \left(i {\partial \!\!\!/ } - m_f \right) Q_-^f(x) 
+ \left\{"+" \leftrightarrow "-" \right\};
\end{align}
for the definition of the Wilson line operator $U[v_\pm(x)]$, see
Eq.~\eqref{eq:6x}.  At leading order in perturbation theory, the
Wilson line operator $U[v_\pm(x)]$ is equal to unity and one finds a
direct coupling between conventional quark fields $\psi$ and reggeized
quark fields $Q_\pm$. The operator $i {\partial \!\!\!/} - m_f$
ensures again vanishing of the coupling for asymptotic
states. Considering the all order Wilson line operator, the induced
Lagrangian turns invariant under local gauge transformations,
Eq.~\eqref{eq:6}.

\section{Working with the high energy effective action}
\label{sec:high-energy-effect-1}

\begin{figure}[htb]
       \centering
   \parbox{.7cm}{\includegraphics[height = 1.8cm]{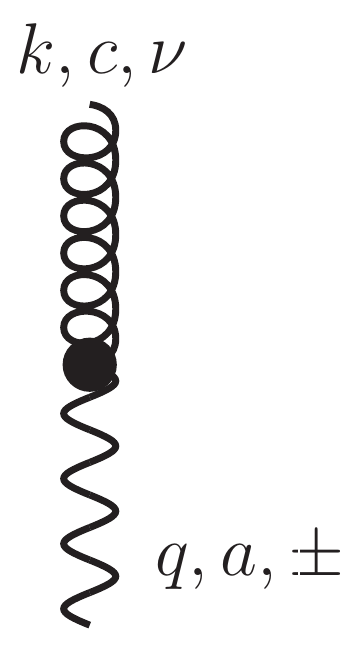}} $=  \displaystyle 
   \begin{array}[h]{ll}
    \\  \\ \frac{- i}{2}{\bf q}^2 \delta^{a c} (n^\pm)^\nu,  \\ \\  \qquad   k^\pm = 0.
   \end{array}  $ 
 \parbox{1.2cm}{ \includegraphics[height = 1.8cm]{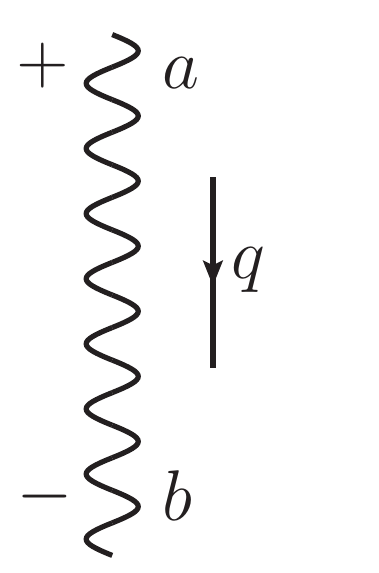}}  $=  \displaystyle    \begin{array}[h]{ll}
    \delta^{ab} \frac{ 2 i}{{\bf q}^2} \end{array}$ 
 \parbox{1.7cm}{\includegraphics[height = 1.8cm]{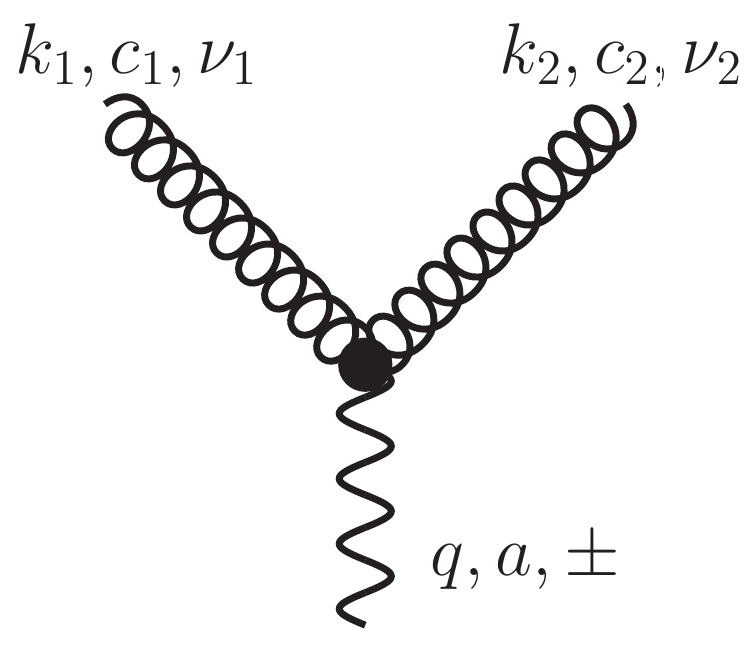}} $ \displaystyle  =  \begin{array}[h]{ll}  \\ \\ \frac{g}{2} f^{c_1 c_2 a} \frac{{\bf q}^2}{k_1^\pm}   (n^\pm)^{\nu_1} (n^\pm)^{\nu_2},  \\ \\ \quad  k_1^\pm  + k_2^\pm  = 0
 \end{array}$
 \\
\parbox{3cm}{\center (a)} \parbox{4cm}{\center (b)} \parbox{5cm}{\center (c)}

\vspace{1cm}
  \parbox{2.4cm}{\includegraphics[height = 1.8cm]{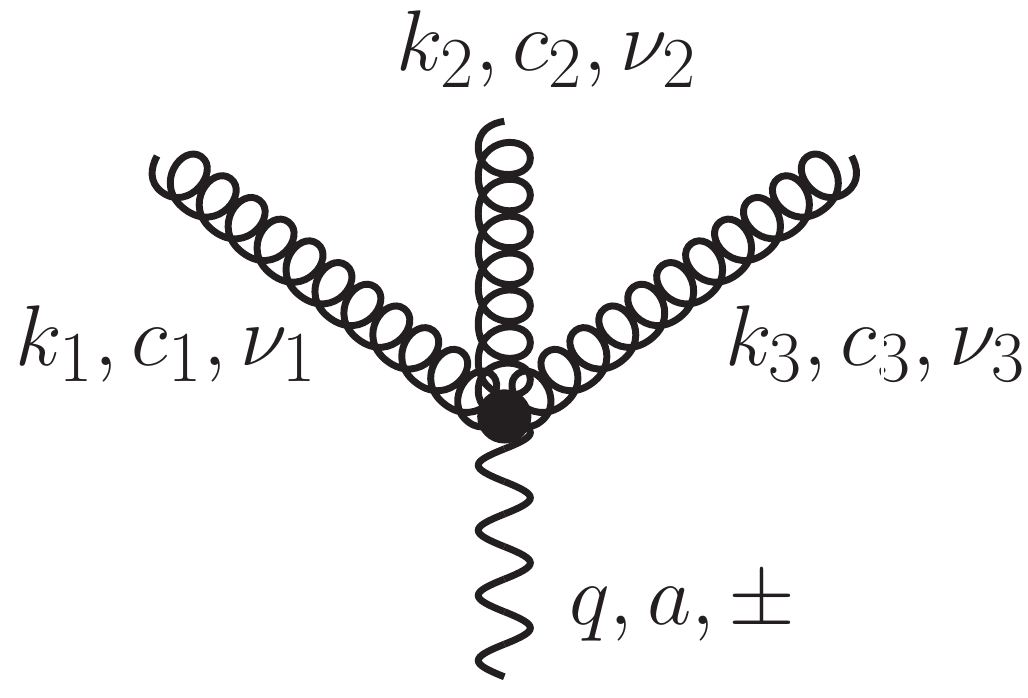}} $ \displaystyle 
   \begin{array}[h]{l}  \displaystyle  \\ \displaystyle= \frac{ig^2}{2} {\bf{q}}^2 
\left(\frac{f^{a_3a_2 e} f^{a_1ea}}{k_3^\pm k_1^\pm} 
+
 \frac{f^{a_3a_1 e} f^{a_2ea}}{k_3^\pm k_2^\pm}\right) (n^\pm)^{\nu_1} (n^\pm)^{\nu_2} (n^\pm)^{\nu_3}, \\ \\
\qquad \qquad   k_1^\pm + k_2^\pm + k_3^\pm = 0.
   \end{array}
$ \\ 
\vspace{.3cm}
\parbox{1cm}{(d)}

 \caption{ Feynman rules for the lowest-order effective vertices
   of the effective action. Wavy lines denote reggeized fields and
   curly lines gluons. Note that in comparison with the Feynman rules
   used in  \cite{Hentschinski:2011tz,Chachamis:2012cc,Chachamis:2012gh, Chachamis:2013hma} we absorbed a factor
   $1/2$ into the vertices which is compensated by changing the
   residue of the reggeized gluon propagator from $1/2$ to $2$.}
\label{fig:3}
\end{figure}

The determination of scattering amplitudes requires at first the
determination of Feynman rules, which has been pioneered in
\cite{Antonov:2004hh}. The Feynman rules of the effective action
Eq.~\eqref{eq:2} comprise in addition to the usual QCD Feynman rules,
the propagator of the reggeized gluon and an infinite number of
so-called induced vertices.  The leading order induced vertices and
propagators are summarized in Fig.~\ref{fig:3}.
In general it is possible to avoid the presence of a direct transition
vertex between gluon and reggeized gluon field. This is most easily
achieved through a shift $v_\pm \to v_\pm + A_\pm$ in the effective
action which removes this transition vertex and gives a direct
coupling of the reggeized gluon field as one of the external fields of
conventional QCD vertices. We will comment on this option further
below. Working with the direct transition vertex, the reggeized gluon
propagator receives also a correction from such a direct transition,
which is already included in the above reggeized gluon propagator.  A
similar set of Feynman rules can be derived for the quark high energy
effective action. In the following we will focus on the high energy
effective action for reggeized gluon exchange, while the bulk of
observations equally applies to the high energy effective action for
reggeized quarks. For a very nice recent treatment of the high energy
effective action for quarks see \cite{Nefedov:2017qzc,Nefedov:2019mrg}.
\\

To start we note that the set of Feynman rules arising from the high
energy effective actions Eqs.~\eqref{eq:2} and \eqref{eq:10} cannot be
used in a straightforward manner. This is immediately clear from the
particular form of the effective action: They contain both the
complete QCD Lagrangian and on top of it the coupling the auxiliary
fields $A_\pm$. The solution to this problem is naturally resolved
through the crucial observation that all the fields inside the high
energy effective action are to be understood as to be restricted to a
certain region local in rapidity:  No strong ordering in light-cone
momenta occurs within such a sector; only the entire
set of light-cone momenta of a certain sector is strongly ordered with
respect to other sectors. Interaction between those sectors takes
place through reggeized gluon exchange. A closely related problem
arises due to the manifestation of so-called rapidity divergences in
certain correlators, which are not regularized by standard methods
such as {\it e.g.} dimensional regularization. Last but not least it
is necessary to provide a suitable pole-description to the light-cone
denominators, which appear in the induced vertices Fig.~\ref{fig:3} to
define properly integrations over loop momenta.  We will shortly
review these problems in the following and discuss commonly solutions
to such problems employed in the literature.

\subsection{Locality in rapidity and subtraction}
\label{sec:locality-rapidity}

\begin{figure}[t]
  \centering
  \begin{tabular}{lcl}
  \parbox{3.3cm}{\includegraphics[width=3.3cm]{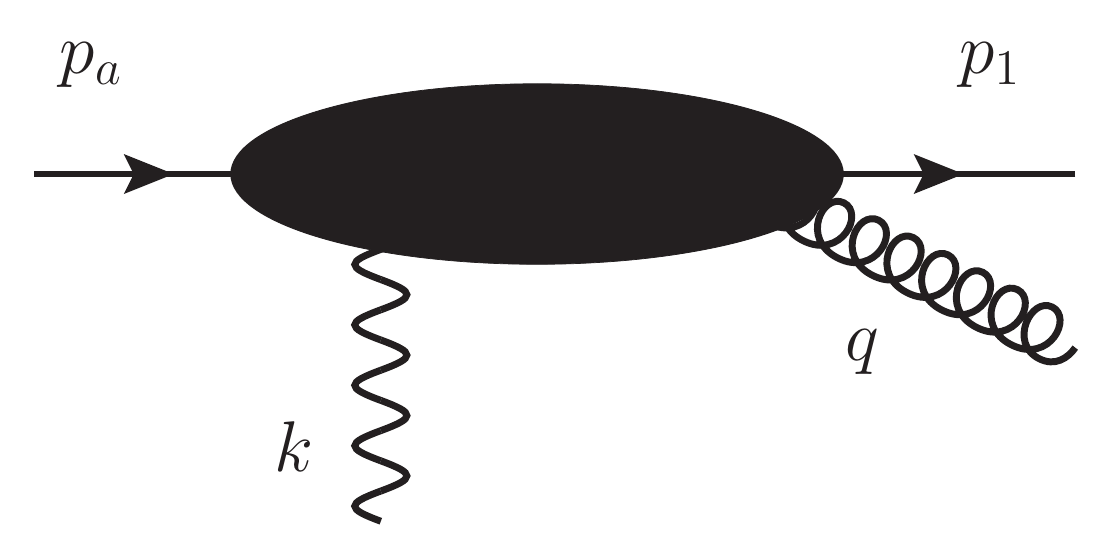}} &
   \parbox{.5cm}{\center $=$} & 
\parbox{3.3cm}{\includegraphics[width=3.3cm]{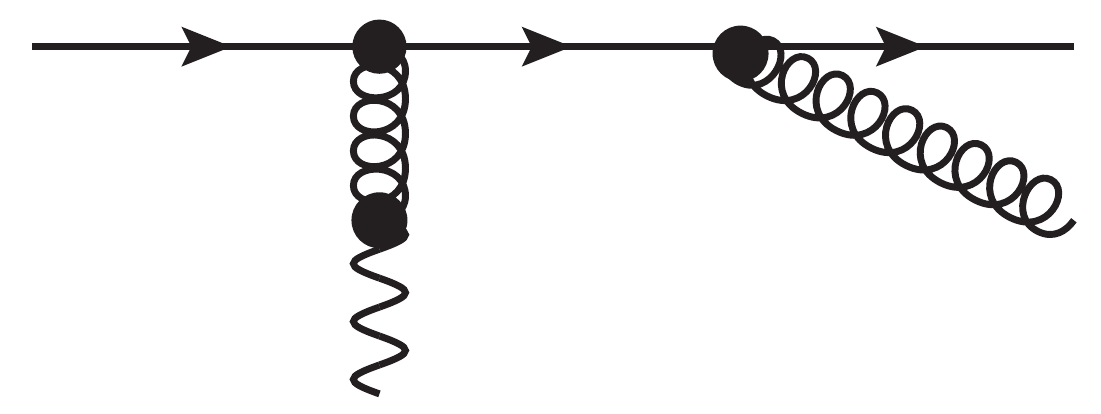}}   \parbox{.5cm}{\center $+$}
\parbox{3.3cm}{\includegraphics[width=3.3cm]{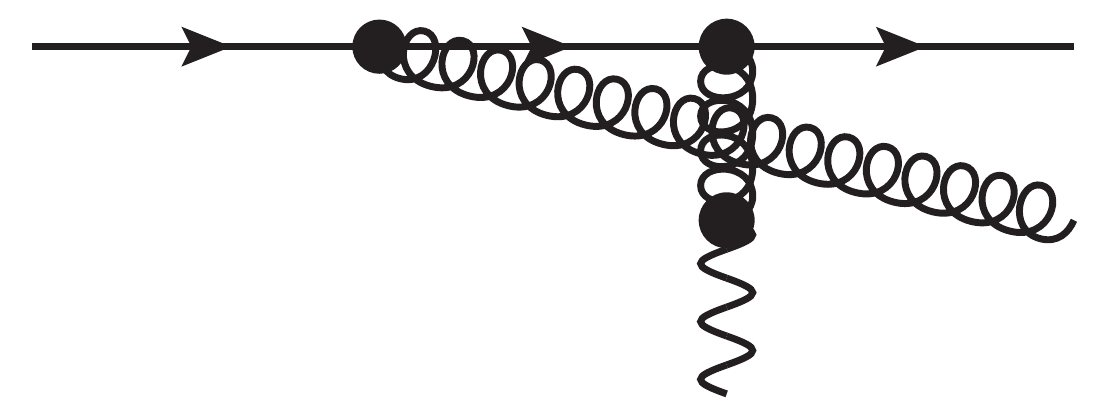}}  \\
& \parbox{.5cm}{\center $+$} &
\parbox{3.3cm}{\includegraphics[width=3.3cm]{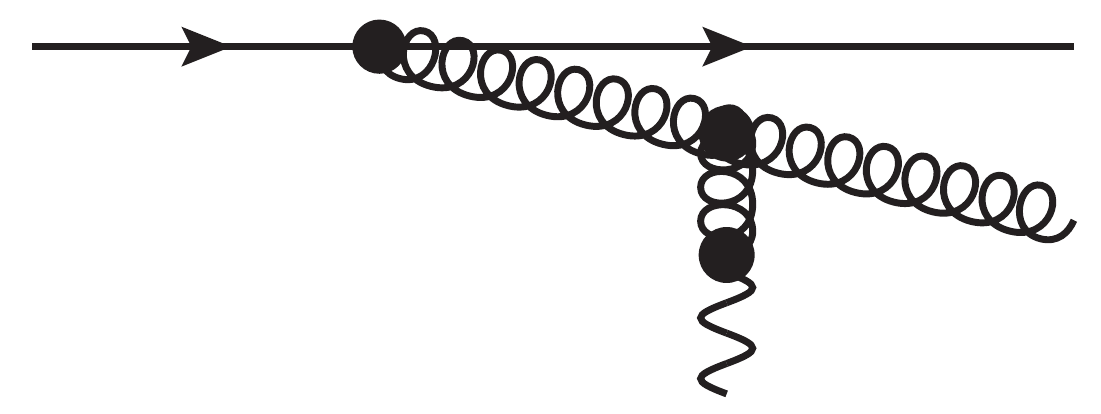}} 
\end{tabular}
  \caption{effective action diagrams for the process quark + reggeized
    gluon $\to$ quark + gluon}
  \label{fig:qqr}
\end{figure}

\begin{figure}[ht]
  \centering
  \parbox{3.5cm}{\includegraphics[width=3.5cm]{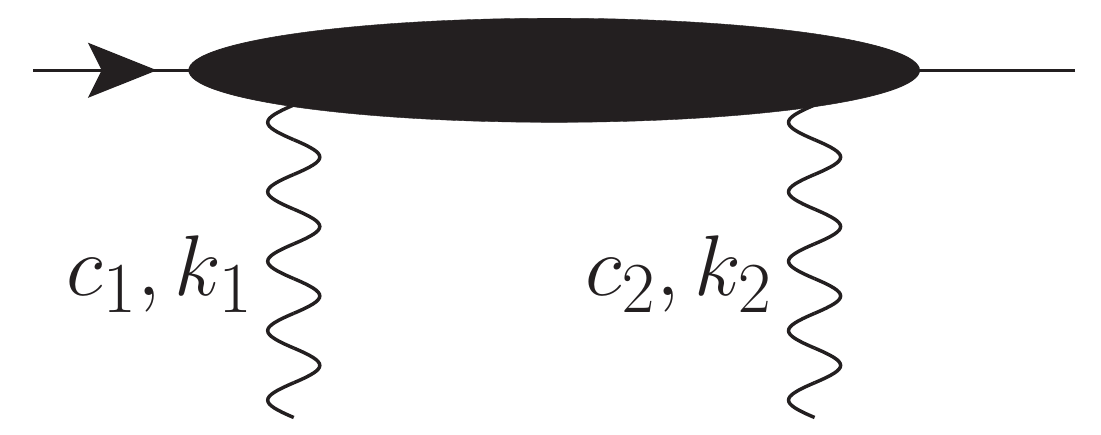}} $=$
  \parbox{3.5cm}{\includegraphics[width=3.5cm]{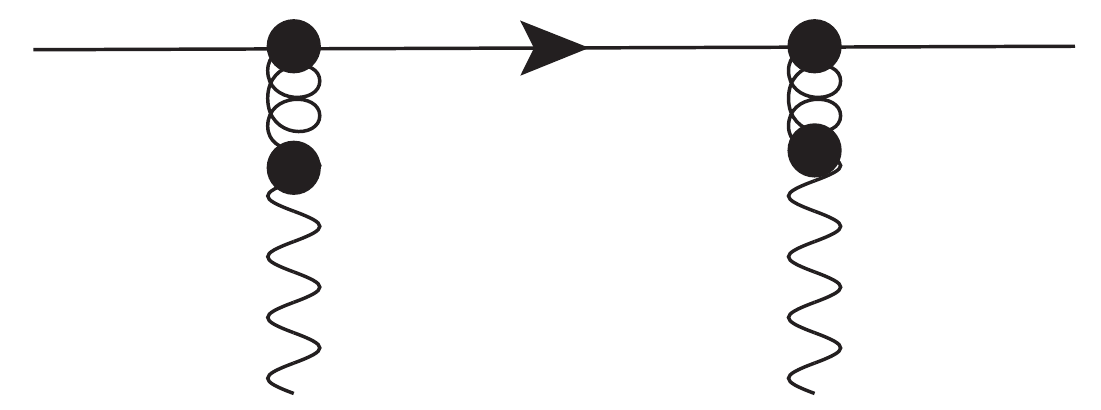}} $+$
  \parbox{3.5cm}{\includegraphics[width=3.5cm]{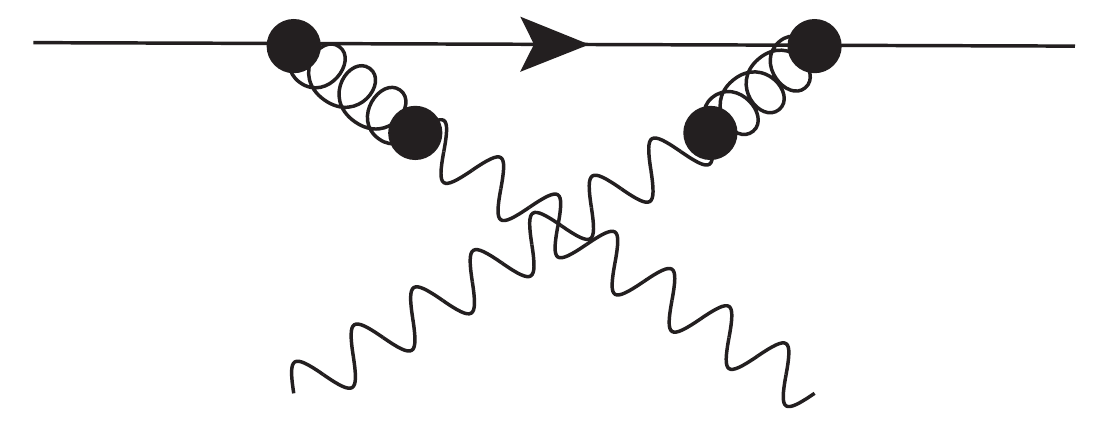}} \\  
\vspace{.7cm}

  \parbox{3.5cm}{\includegraphics[width=3.5cm]{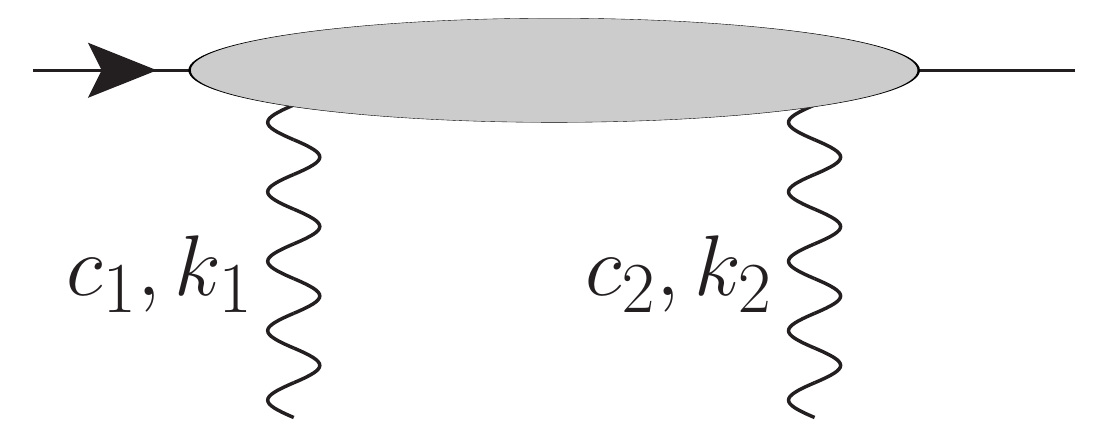}} 
 $=$ \parbox{3.5cm}{\includegraphics[width=3.5cm]{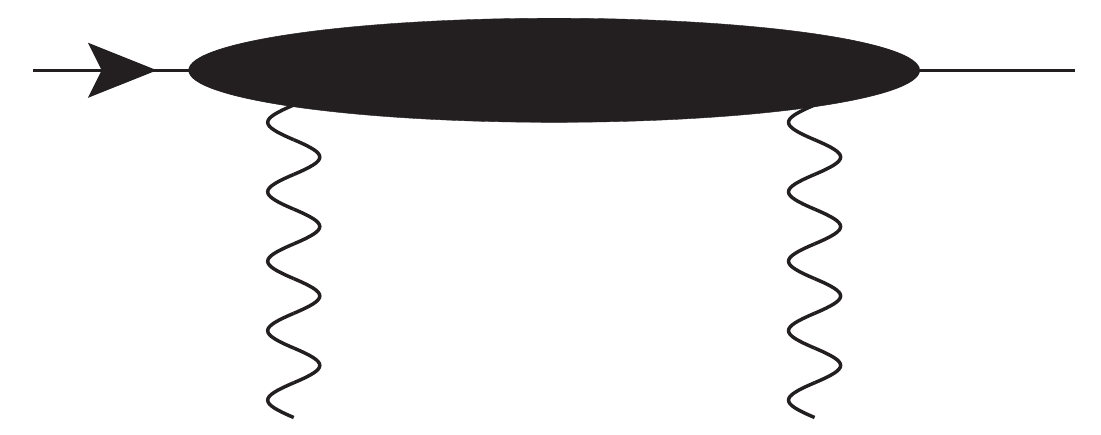}}
    $-$ \parbox{3.5cm}{\includegraphics[width=3.5cm]{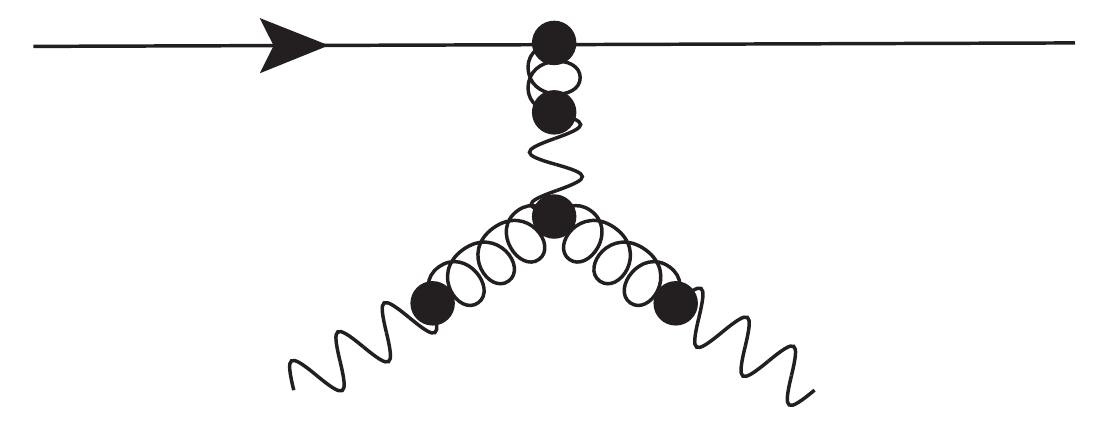}}
  \caption{First line: Amplitude for 2 reggeized gluons coupling to a
    quark. Second line: Amplitude with the factorized contribution subtracted.}
  \label{fig:q_rr}
\end{figure}

The problem is most easily illustrated using a particular example: We
consider the additional emission of a gluon in the the interaction of
a quark with a reggeized gluon. For full details and explicit
expressions we refer  to \cite{Hentschinski:2011tz, Chachamis:2012mw}.
The scattering amplitude can be obtained from the high energy
effective action through evaluating the sets of diagrams, shown in
Fig.~\ref{fig:qqr}
 The resulting scattering amplitude  does not vanish in the strongly ordered 
region  $p^+ \gg q^+$; it  is therefore  non-local in rapidity. 
\begin{figure}[t]
  \centering
  \parbox{1.8cm}{\center \includegraphics[height=3cm]{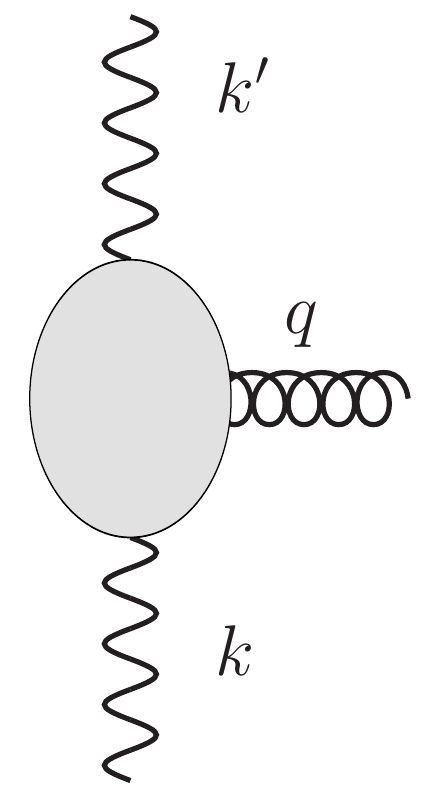}} $=$
 \parbox{1.8cm}{\center \includegraphics[height=3cm]{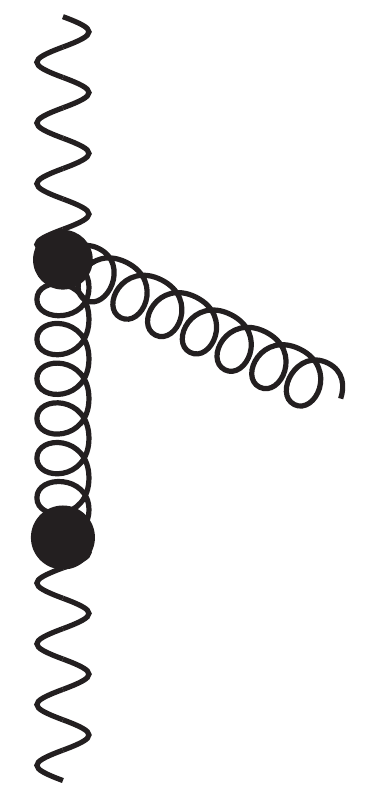}} $+$
 \parbox{1.8cm}{\center \includegraphics[height=3cm]{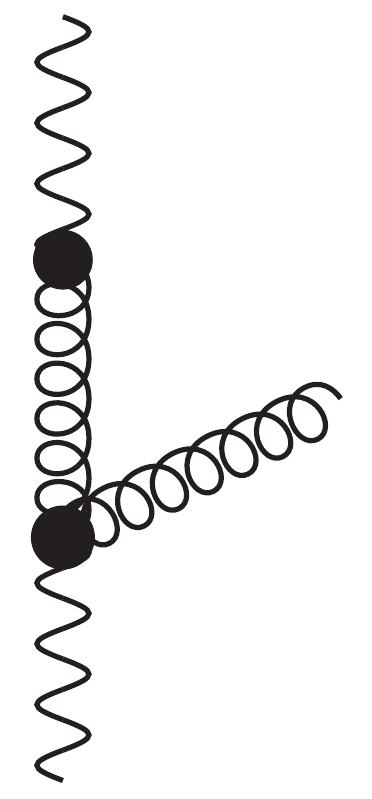}} $+$
 \parbox{1.8cm}{\center \includegraphics[height=3cm]{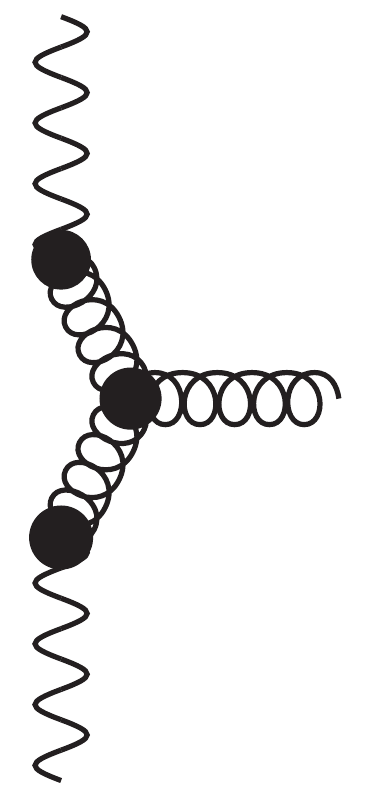}} 
  \caption{The central gluon production vertex, {\it i.e.} the Lipatov vertex, in terms of  effective
    action diagrams}
  \label{fig:nonlocal_production}
\end{figure}
Apart from the above matrix element Fig.~\ref{fig:qqr}, the effective
action allows also for the construction of a scattering amplitude for
the same process, which combines two different sectors, connected
through an internal reggeized gluon line: the 2 sectors are in this
case given by the coupling of a quark to a reggeized gluon and the
emissions of a gluon from a reggeized gluon, the famous gluon
production or Lipatov vertex, see \cite{Antonov:2004hh} for a
derivation of this vertex from the effective action. Analyzing the
explicit expressions corresponding to Fig.~\ref{fig:qqr}, it is
evident that Fig.~\ref{fig:qqr} turns into Fig.~\ref{fig:q_rr} in the
strongly ordered region $p^+ \gg q^+$, where gluon and quark are
widely separated in rapidity. In this particular example, it would be
straightforward to include an explicit cut-off on the final state that
separates Fig.~\ref{fig:qqr} from Fig.~\ref{fig:nonlocal_production}.
With increasing complexity of the final states and/or the case of
higher-loop corrections, this becomes rather cumbersome. It is
therefore more suitable, to simple \emph{subtract} diagrams with an
\emph{internal} reggeized gluon line from the regarding correlator,
removing in this way the contribution which causes the non-locality in
rapidity. For the above example one finds that the combination
\begin{align}
  \label{eq:subtractQQR}
   \parbox{3.8cm}{\includegraphics[height=1.5cm]{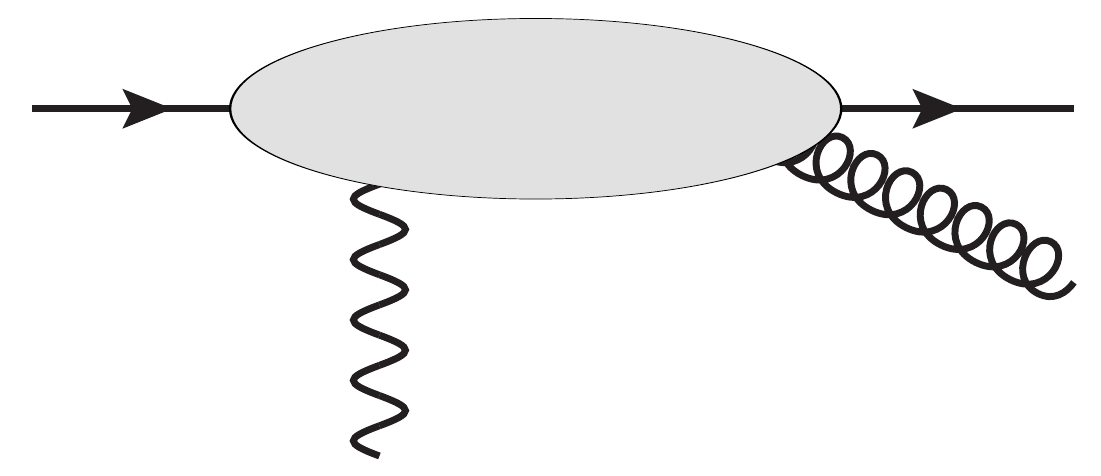}}  
&  =  \parbox{3.8cm}{\includegraphics[height=1.5cm]{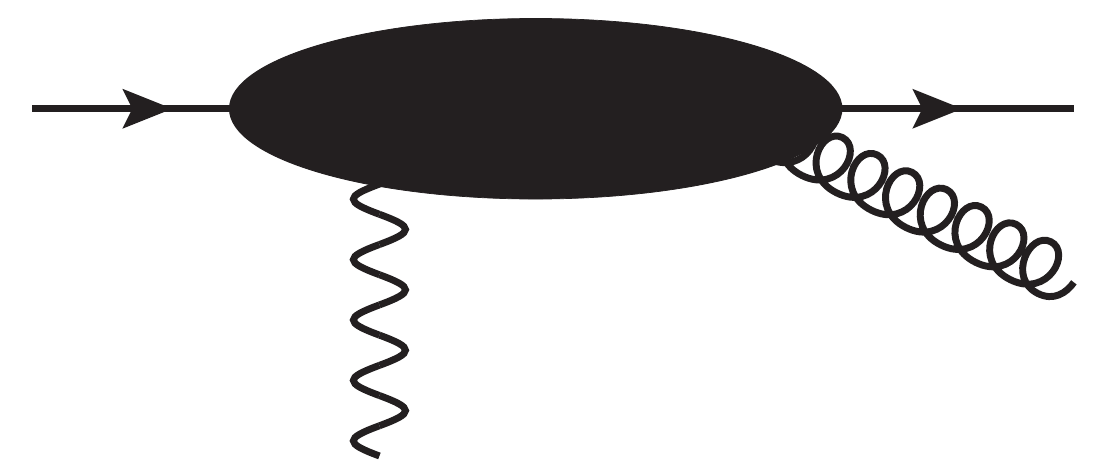}}
 -   \parbox{3.8cm}{\includegraphics[height=1.5cm]{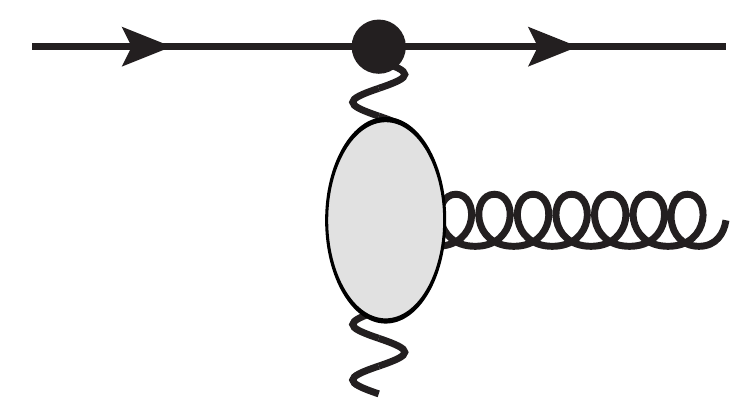}}
\end{align}
is local in rapidity, in the sense that the resulting expression is not growing in the strongly ordered region  $p^+ \gg q^+$ but instead assumes a constant value. \\

In general, the following set of rules can be formulated for the
determination of a (sub-) amplitude, local in rapidity within the
effective action, using perturbation theory:
\begin{itemize}
\item[a)] determine the correlator of interest from the high energy
  effective action, to the desired order in perturbation theory, with
  the reggeized gluon field treated as a background field
\item[b)] subtract disconnected contributions which contain internal
  reggeized gluon lines (which are already themselves subtracted) to
  the desired order in perturbation theory.
\end{itemize}
A similar situation arises for diagrams with multiple reggeized gluon
exchange, such as Fig.~\ref{fig:q_rr}, first line.
In the limit where the squared center-of-mass energy
$s = (p_a + k_1)^2$ tends to infinity, the contribution diverges
logarithmically $~ 1/k_1^-$.  Similar to the previous example, the
high energy limit of this correlator is captured by the corresponding
diagram with an internal reggeized gluon exchange, such that the
subtracted contribution, Fig.~\ref{fig:q_rr}, second line, is local in
rapidity in the sense that the expression vanishes up to terms
$\mathcal{O}(1/(k_1^-)^2$ in the high energy limit
\cite{Hentschinski:2009zz}, see also the discussion of a related
observation in \cite{Braun:2019gjy}.

\subsection{Pole prescription for induced vertices}
\label{sec:pole-prescr-induc}

At first sight there is no need to chose a particular pole
prescription: the poles in $1/\partial_\pm$ arise from propagators of
particles with significantly different rapidity, which emit gluons into
the rapidity cluster under consideration. The corresponding particle
virtualities $~ k^\pm$ are therefore large. Based on this argument,
the pole prescription of these denominators can be ignored and one
arrives directly at the induced vertices Fig.~\ref{fig:3}. Most
notably, the color structure of these vertices is given structure
constants $f^{abc}$ only, which arise from commutators of SU$(N_c)$
generators of the fields $v(x) = -i t^a v^a(x)$ in the functional
$T_\pm(v_\pm)$. Since the color structure of induced vertices arises
from commutators of SU$(N_c)$ generators, the color structure is
independent of the representation of the fields $v(x)$. The resulting
induced vertices are therefore universal. They do not depend on the
color representation of the specific particle from which the gluons
have been emitted.
\\

Even though the overall result must be independent of the pole
prescription due to the above arguments, it is nevertheless necessary
to chose a certain prescription in actual calculations. To have
consistency for the evaluation of different elements (which eventually
one needs to combine) it is highly recommendable to do so at the level
of the effective Lagrangian or equivalently the resulting Feynman
rules.  At first sight the problem appears to be well defined:
Requiring convergence of the path ordered exponential Eq.~\eqref{eq:6}
at $-\infty$ naturally defines a pole prescription of the light-cone
denominator: it induces the replacement
$\partial_\pm \to \partial_\pm + \epsilon $. There are various reasons
why this is not the generally accepted treatment (see in particular
the discussion in \cite{Lipatov:2016ged}): At the very first, using
such a pole prescription, the effective Lagrangian is no longer
hermitian and unitary of theory is formally lost (which itself is an
essential requirement for any effective field theory description to
study QCD high energy limit). In practice this problem can be overcome
by replacing the path ordered exponential Eq.~\eqref{eq:6} by the
combination
\begin{align}
  \label{eq:17}
  U[v_\pm(x)] \to \frac{1}{2} \left( U[v_\pm(x)] +  U^\dagger[v_\pm(x)] \right).
\end{align}
A similar prescription has been adapted in the original formulation of
the quark effective action \cite{Lipatov:2000se}; for a related
discussion in the case of the effective action for gluons see
\cite{Nefedov:2017qzc}. For the induced vertex Fig.~\ref{fig:3}.c,
such a prescription leads to a principal value prescription, {\it
  i.e.}
\begin{align}
  \label{eq:18}
  \frac{1}{k_1^\pm} & \to \frac{1}{2} \left(  \frac{1}{k_1^\pm + i \epsilon} +  \frac{1}{k_1^\pm - i \epsilon} \right) = \frac{k_1^\pm}{(k_1^\pm)^2 + \epsilon^2}.
\end{align}
Such a  prescription is not only attractive from the point of view of
hermicity of the high energy effective action, it further provides
vertices which are odd under $k_1^\pm \to -k_1^\pm$ -- a property in
accordance with negative signature of reggeized gluon (which
eventually arises from the reggeized gluon fields). While this pole
prescription provides a simple and straightforward definition of the
pole prescription and appears suitable for the use in calculations, it
gives rise to certain symmetric color structures which cannot be
expressed as commutators (and therefore enhances the color structure of the induced vertices in Fig.~\ref{fig:3}). Since symmetric color tensors depend in
general on the representation of the SU$(N_c)$ generators,
universality of these vertices is therefore  lost\footnote{Applying a
  consistent subtraction mechanism, the dependence on such color
  tensors must drop in the final result, {\it i.e.} on the level of
  the observable. It should be therefore possible to carry out actual
  higher order calculations within this prescription}.  To avoid this
problem, in \cite{Hentschinski:2011xg} a perturbative prescription has
been developed which projects out order by order in perturbation
theory symmetric color structures.  The resulting pole prescription
respects then Bose symmetry and universality of the induced vertices.  For the $\mathcal{O}(g)$ vertex, see 
Fig.~\ref{fig:3}.c, this corresponds to replacing the denominator
$1/k_1^\pm$ by a Cauchy principal value. For the $\mathcal{O}(g^2)$
vertex, see Fig.~\ref{fig:3}.c, the combinations of denominators $1/(k_3^\pm
k_1^\pm)$ and $1/(k_3^\pm k_2^\pm)$ are replaced by functions $g_2^\pm
(3,2,1)$ and $g_2^\pm (3,1,2)$ respectively with
\begin{align}
  \label{eq:cpv_rep}
g_2^\pm(i,j,m) = 
    \bigg[&  \frac{1}{[k_i^\pm][k_m^\pm]} + \frac{\pi^2}{3}\delta(k_i^\pm)\delta(k_m^\pm) \bigg],
& \frac{1}{[k]} & \equiv \frac{1}{2} \left(\frac{1}{k + i 0} + \frac{1}{k - i0} \right).
\end{align}
Similar functions can be constructed for the higher order induced
vertices. One of the main advantages over simply defining all
denominators as principal values is that the function $g_2$ (and
corresponding higher order versions) satisfy eikonal identities in the
sense of the theory of distributions {\it i.e.}
\begin{align}
  \label{eq:20}
  g_2^{\pm} (1,2,3) +  g_2^{\pm} (2,3,1) +  g_2^{\pm} (3,1,2) & = 0.
\end{align}
Finally we would like to highlight that Lipatov himself proposed a certain pole-prescription in \cite{Lipatov:2016ged}
\begin{align}
  \label{eq:19}
  T_\pm[v_\pm] & = \frac{1}{2}\mathcal{P}  
\frac{ e^{-\frac{g}{4} \int_{-\infty}^{x_\pm} dx^{\pm'} v_\pm}}{ e^{
-\frac{g}{4} \int^{\infty}_{x_\pm} dx^{\pm'} v_\pm}} \left[
\mathcal{P} e^{\frac{g}{4} \int_{-\infty}^{x^\pm} dx^{\pm'} v_\pm}
+
\bar{\mathcal{P}} e^{-\frac{g}{4} \int_{-\infty}^{x^\pm} dx^{\pm'} v_\pm}
\right],
\end{align}
where ${\mathcal{P}}$ ($\bar{\mathcal{P}}$) denotes  (anti-)path ordering. To the best our knowledge the vertices resulting from this particular pole prescription have not been explored so far. 

\subsection{Regularization of high energy singularities}
\label{sec:regul-high-energy}

Last but not least it is necessary to address the problem that higher
order corrections lead to so-called rapidity divergences which are not
regularized by conventional methods, such as dimensional
regularization. A very nice review of different regulators which can
be used can be found in \cite{Nefedov:2019mrg}. A very convenient
choice is given by tilting the light-cone direction of the effective
action slightly against the light cone through introducing a regulator
$\rho$, evaluated in the limit $\rho \to \infty$
\begin{align}
  \label{eq:deform}
  n^-  & \to n_a = e^{-\rho} n^+ + n^-, \notag \\
  n^+ & \to n_b = n^+ + e^{-\rho} n^-.
\end{align}
This regulator has the important advantage over other regulators (such as regulation by a hard cut-off or analytic regulations) since it respects gauge invariance properties of the high energy effective action.  

\section{The gluon Regge trajectory and  the effective action}
\label{sec:gluontrajectory-effectiveaction}
As an example of an explicit higher order calculations which combines
all of the above aspects, we review in the following the determination
of the gluon Regge trajectory from the high energy effective action,
see \cite{Chachamis:2012gh,Chachamis:2013hma} for full details. This calculation requires
\begin{itemize}
\item determination of  the propagator of the reggeized gluon to the desired order in perturbation theory;
\item a transition function for  its rapidity divergences.
\end{itemize}
As a start it is necessary to to determine the self-energies of the
reggeized gluon. Within the subtraction procedure
Sec.~\ref{sec:locality-rapidity}, this requires
\begin{itemize}
\item[(a)]
 determination of the self-energy of the reggeized gluon from the effective action, with the reggeized gluon treated as a background field;
\item[(b)] 
subtraction of all  disconnected contributions which contain internal reggeized gluon lines.
\end{itemize}
Using the pole prescription of \cite{Hentschinski:2011xg}, at 1-loop,
all diagrams with internal reggeized gluon lines vanish, and no
subtraction occurs; the contributing diagrams are
\begin{align}
   \Sigma^{(1)}\left(\rho; \epsilon, \frac{{\bf q}^2}{\mu^2}    \right)   & = 
  \parbox{1cm}{\vspace{0.1cm} \includegraphics[height = 1.8cm]{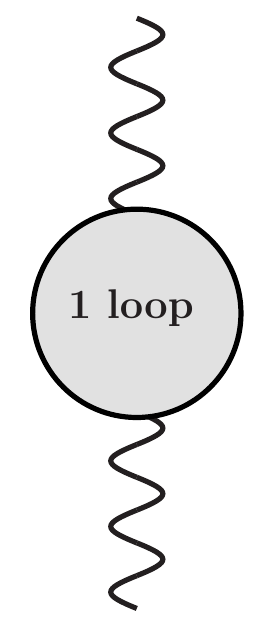}}
= 
    \parbox{.7cm}{\vspace{0.1cm}  \includegraphics[height = 1.8cm]{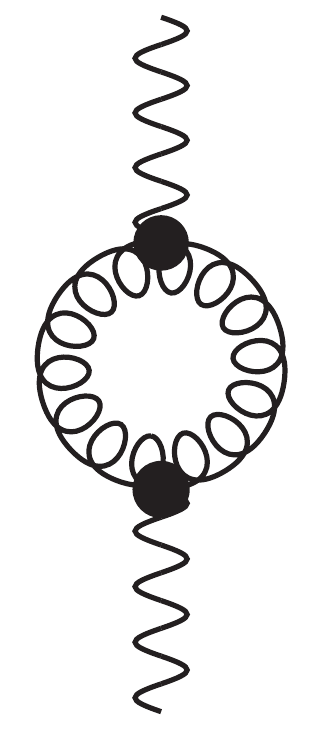}}
  + 
  \parbox{.7cm}{\vspace{0.1cm} \includegraphics[height = 1.8cm]{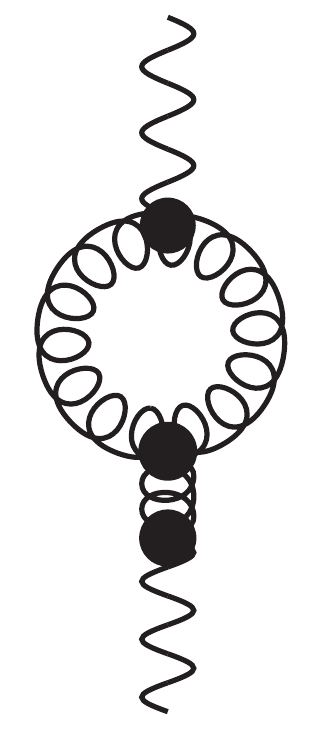}} 
 +
  \parbox{.7cm}{\vspace{0.1cm} \includegraphics[height = 1.8cm]{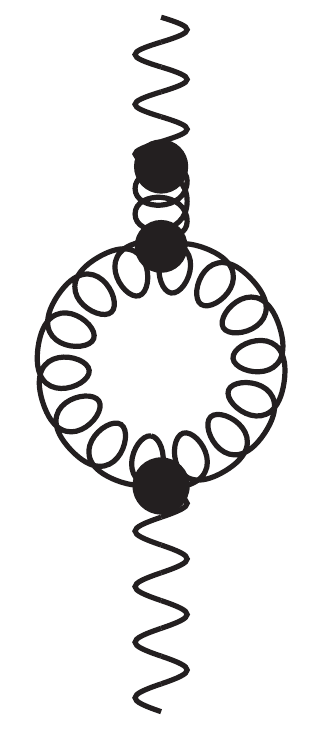}} 
 +
  \parbox{.7cm}{\vspace{0.1cm} \includegraphics[height = 1.8cm]{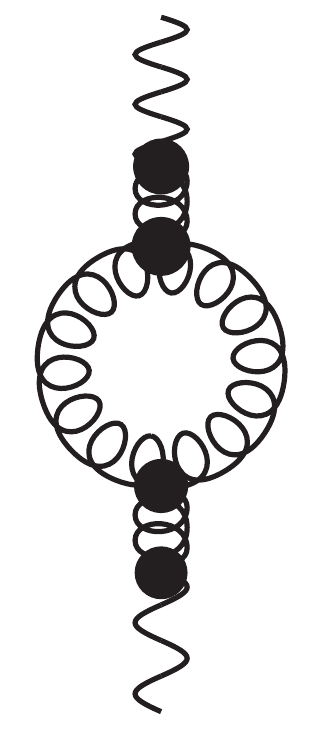}} 
 +
  \parbox{.7cm}{\vspace{0.1cm} \includegraphics[height = 1.8cm]{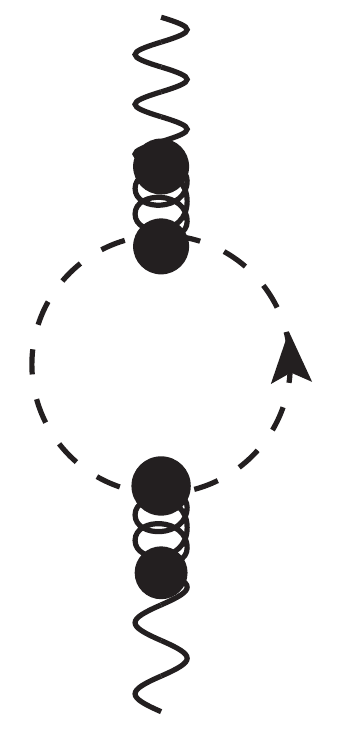}}
+
  \parbox{.7cm}{\vspace{0.1cm} \includegraphics[height = 1.8cm]{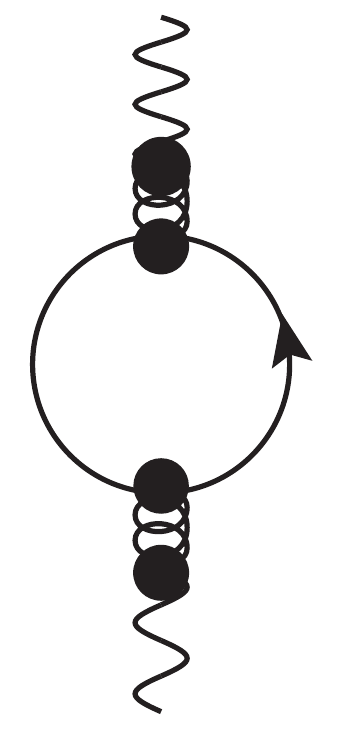}}\, \,\,.
\label{eq:self_1loop}
\end{align}
\\
The first of the above diagrams is divergent and requires
regularization. To achieve a consistent treatment, we tilt the
light-cone direction for all diagrams and take the limit  $\rho  \to
\infty$  after the evaluation of integrals. Keeping only finite and
divergent terms, one finds in $d = 4 + 2\epsilon$
dimensions\footnote{In the original result present in \cite{Hentschinski:2011tz}
  and reproduced in \cite{Chachamis:2012mw}, a finite result for the second and
  third diagram has been erroneously included, see also \cite{Chachamis:2012cc, Chachamis:2012gh, Chachamis:2013hma}}:
\begin{align}
\label{eq:self_1loopRES}
   \frac{\Sigma^{(1)}\left(\rho; \epsilon, \frac{{\bf q}^2}{\mu^2}    \right) }{(-2i {\bf q}^2) }     
 &=
\bar{g}^2 c_\Gamma \left(\frac{{\bf q}^2}{\mu^2} \right)^\epsilon  
  \bigg\{  \frac{ i\pi - 2 \rho}{\epsilon}         
- \frac{   5 + 3 \epsilon - \frac{n_f}{N_c} (2 + 2\epsilon)}{(1 + 2 \epsilon)(3 + 2\epsilon)\epsilon} \bigg\},
\end{align}
and
\begin{align}
  \label{eq:gbar}
  \bar{g}^2 & =  \frac{g^2 N_c \Gamma(1 - \epsilon)}{(4 \pi)^{2 + \epsilon}}, & c_\Gamma & = \frac{ \Gamma^2(1 + \epsilon)}{\Gamma(1 + 2 \epsilon)}.
\end{align}
To determine the 2-loop self energy it is now needed to subtract
disconnected diagrams, whereas diagrams with multiple internal
reggeized gluons can be shown to yield zero result. Schematically one
has
\begin{align}
  \label{eq:coeff_2loop}
  \Sigma^{(2)}\left(\rho; \epsilon, \frac{{\bf q}^2}{\mu^2}    \right)   & =  
   \parbox{2cm}{\center \includegraphics[height = 2.5cm]{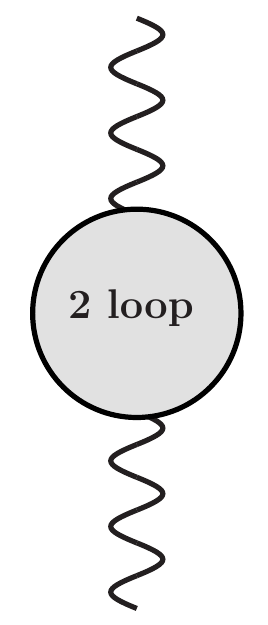}}  
  = 
  \parbox{2cm}{\center \includegraphics[height = 2.5cm]{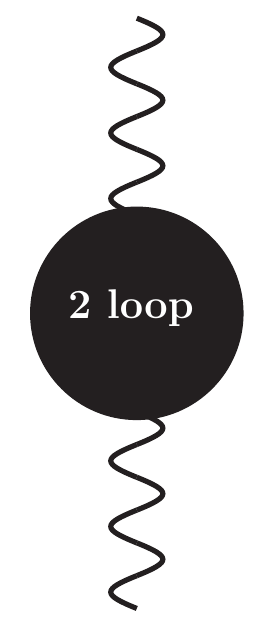}}
  -
  \parbox{2cm}{\center \includegraphics[height = 2.5cm]{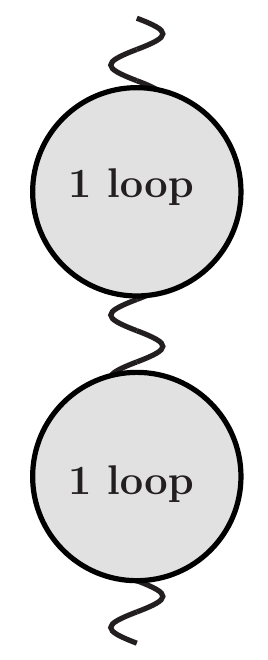}}.
\end{align}
The black blob denotes the unsubtracted 2-loop reggeized gluon
self-energy. It  is obtained through the direct application of the
Feynman rules of the effective action (with the reggeized gluon itself
treated as a background field). The resulting 2-loop integrals have been determined in~\cite{Chachamis:2012gh, Chachamis:2013hma}. The result for
$n_f$ flavor reads
\begin{align}
\label{meister}
& \Sigma^{(2)}\left(\rho,\frac{\bf{q}^2}{\mu^2}\right)
= (- 2i{\bf q}^2)\frac{g^4 N_c^2}{(4\pi)^4}
\bigg\{
-\bigg[
\frac{2}{\epsilon^2}+\frac{4(1-\tau)}{\epsilon}
+ 4(1-\tau)^2- \frac{\pi^2}{3}
\bigg]\rho^2
\notag \\
&+
\bigg[
 \frac{1}{3\epsilon^2}+\frac{1}{9\epsilon}  +\frac{\pi^2}{3 \epsilon}-\frac{2 \tau}{3 \epsilon}+\frac{\pi^2(11-12\tau)}{18}
 +\frac{16}{27} - \frac{2 }{9} \tau + \frac{2}{3} \tau^2 -2\zeta(3)\bigg)\bigg]\rho \bigg \} 
\notag \\
&
+ \frac{ n_f}{N_c} \left( \frac{2}{3\epsilon} + \frac{n_f (6 - 36 \tau)}{27 \epsilon} + \frac{32 - 3 \pi^2 - 12 \tau + 36 \tau^2}{27}   \right)
+ \mathcal{O}(\epsilon) + \mathcal{O}(\rho^0).&
\end{align}
with $\tau = 1 - \ln \frac{{\bf q}^2 e^{\gamma_E}}{ 4 \pi \mu^2 }$. To
obtain the gluon trajectory, we need to construct next the (bare)
two-loop reggeized gluon propagators
\begin{align}
  \label{eq:barepropR}
   G \left(\rho; \epsilon, {\bf q}^2, \mu^2   \right)
&=
\frac{i/2}{{\bf q}^2} \left\{ 1 + \frac{i/2}{{\bf q}^2} \Sigma \left(\rho; \epsilon, \frac{{\bf q}^2}{\mu^2}    \right)  + \left[  \frac{i/2}{{\bf q}^2} \Sigma \left(\rho; \epsilon, \frac{{\bf q}^2}{\mu^2}   \right)\right] ^2 + \ldots   \right\},
\end{align}
with the reggeized gluon self energy
\begin{align}
  \label{eq:sigma}
   \Sigma \left(\rho; \epsilon, \frac{{\bf q}^2}{\mu^2}   \right) & =   \Sigma^{(1)} \left(\rho; \epsilon, \frac{{\bf q}^2}{\mu^2}   \right) +  \Sigma^{(2)} \left(\rho; \epsilon, \frac{{\bf q}^2}{\mu^2}   \right) + \ldots .
\end{align}
Apparently Eq.~\eqref{eq:self_1loop} is divergent in the limit
$\rho \to \infty$. In \cite{Hentschinski:2011tz,Chachamis:2012cc} it has been
demonstrated by explicit calculations that these divergences cancel at
one-loop level against divergences in the couplings of the reggeized
gluon to external particles. The entire one-loop amplitude is then
found to be free of any high energy singularity in $\rho$.
Consistency of high energy factorization as formulated within the high
energy effective action requires that such a cancellation holds also
beyond one loop. To make this cancelation explicit, we  introduce  transition functions $Z^\pm$.  For
explicit examples we refer the reader to \cite{Hentschinski:2011tz, Chachamis:2012cc}.
In particular we define the  renormalized reggeized gluon propagator as
\begin{align}
\label{eq:renor}
G^{\rm R}(\eta;\epsilon,\bf{q}^2,\mu^2) 
&= \frac{G(\rho;\epsilon,\bf{q}^2,\mu^2)}{Z^+\left(\eta,\rho;\epsilon,\frac{\bf{q}^2}{\mu^2}\right)Z^-\left(\eta,\rho;\epsilon,\frac{\bf{q}^2}{\mu^2}\right)},
\end{align}
 In their most general form these
transition functions are parametrized as
\begin{align}
\label{eq:param}
Z^\pm\left(\eta,\rho;\epsilon,\frac{\bf{q}^2}{\mu^2}\right)=\exp\left[\frac{\rho- \eta}{2}\omega\left(\epsilon,\frac{\bf{q}^2}{\mu^2}\right)   +   f^\pm\left(\epsilon,\frac{\bf{q}^2}{\mu^2}\right)\right].
\end{align}
The coefficient of the $\rho$-divergent term defines the gluon Regge trajectory $\omega(\epsilon, {\bf q}^2)$, 
\begin{align}
  \label{eq:perturexpomega}
  \omega\left(\epsilon, \frac{{\bf q}^2}{\mu^2} \right) &= 
 \omega^{(1)}\left(\epsilon, \frac{{\bf q}^2}{\mu^2} \right)  
+
 \omega^{(2)}\left(\epsilon, \frac{{\bf q}^2}{\mu^2} \right)  + \ldots,
\end{align}
which is determined by the requirement that the renormalized reggeized
gluon propagator must be free of high energy divergences, {\it i.e.}
$\rho$ independent.  At one loop one obtains
\begin{align}
  \label{eq:omega1}
  \omega^{(1)}\left(\epsilon, \frac{{\bf q}^2}{\mu^2} \right) & = -\frac{2 \bar{g}^2 \Gamma^2(1 + \epsilon)}{\Gamma(1 + 2 \epsilon)\epsilon } \left(\frac{{\bf q}^2}{\mu^2} \right)^\epsilon.  
\end{align}
The function $f^\pm(\epsilon, {\bf q}^2)$ parametrizes finite
contributions and is, in principle, arbitrary. Symmetry of the
scattering amplitude requires $f^+ = f^- = f$, Regge theory suggests
fixing it in such  that terms which are not enhanced in $\rho$
are entirely transferred from the reggeized gluon propagators to the vertices, to which the reggeized gluon
couples. With 
\begin{align}
  \label{eq:expansionofF}
  f\left(\epsilon, \frac{{\bf q}^2}{\mu^2} \right) & = f^{(1)}\left(\epsilon, \frac{{\bf q}^2}{\mu^2} \right) + f^{(2)}\left(\epsilon, \frac{{\bf q}^2}{\mu^2} \right) \ldots
\end{align}
we obtain from  Eq.~\eqref{eq:self_1loop}
\begin{align}
  \label{eq:f1loop}
   f^{(1)}\left(\epsilon, \frac{{\bf q}^2}{\mu^2} \right) & =  \frac{ \bar{g}^2 \Gamma^2(1 + \epsilon)}{\Gamma(1 + 2 \epsilon)} \left(\frac{{\bf q}^2}{\mu^2} \right)^\epsilon  
         \frac{(-1)}{(1 + 2 \epsilon)2 \epsilon} \bigg[    \frac{5 + 3\epsilon}{3 + 2 \epsilon} 
-\frac{n_f}{N_c} \left( \frac{2 + 2\epsilon}{3 + 2\epsilon} \right)\bigg]  ,
\end{align}
and consequently
\begin{align}
  \label{eq:omega2_defined}
 \omega^{(2)}\left(\epsilon, \frac{{\bf q}^2}{\mu^2} \right)  & = 
 \lim_{\rho \to \infty} \frac{1}{\rho} \bigg[ \frac{\Sigma^{(2)}}{(-2i{\bf q}^2)}  + \frac{\rho^2}{2} \left(\omega^{(1)}\right)^2 + 2 \rho   f^{(1)} \omega^{(1)}  \bigg],
\end{align}
where we omitted at the right hand side the dependencies on $\epsilon$
and ${\bf q}^2/\mu^2$ and  expanded
$\Sigma^{(1)}$ in terms of the functions $\omega^{(1)}$ and
$f^{(1)}$.  We obtain
\begin{equation}
\omega^{(2)}({ \bf q}^2)=\frac{(\omega^{(1)}({\bf q}^2))^2}{4}
\left[
\frac{11}{3} - \frac{2 n_f}{3 N_c} +\left(\frac{\pi^2}{3}-\frac{67}{9}\right)\epsilon+\left(\frac{404}{27}-2\zeta(3)\right)\epsilon^2
\right],
\end{equation}
in agreement with the results in the literature\cite{Fadin:1996tb}. The parameters $\eta$ in the transition functions  are arbitrary; their role is analogous to
the renormalization scale in UV renormalization or the factorization
scale in collinear factorization. It gives rise to a  dependence of the reggeized gluon propagator on the the factorization parameter  $\eta$ from which a renormalization group equation  (RGE) results:
\begin{align}
  \label{eq:21}
  \frac{d}{d \eta} G^R\left(\eta; \epsilon, \frac{{\bf q}^2}{\mu^2}\right) & = \omega\left(\epsilon, \frac{{\bf q}^2}{\mu^2}\right ) G^R\left(\eta; \epsilon, \frac{{\bf q}^2}{\mu^2}\right).
\end{align}
With the above choice for the function $f^\pm$, one has
\begin{align}
  \label{eq:22}
   G^R\left(0; \epsilon, \frac{{\bf q}^2}{\mu^2}\right) & = \frac{i/2}{{\bf q}^2},
\end{align}
and
\begin{align}
  \label{eq:23}
   G^R\left(\eta; \epsilon, \frac{{\bf q}^2}{\mu^2}\right) & = \frac{i/2}{{\bf q}^2} \exp\left[ \eta  \cdot \omega\left(\epsilon, \frac{{\bf q}^2}{\mu^2}\right ) \right].
\end{align}
The value of the parameter of $\eta$ is to be fixed by the impact factors which describe coupling of the reggeized gluon to external particles. In the case of partonic scattering, the natural choice is given by $\eta = \ln s/|t|$ with $t= - {\bf q}^2$ and $s$ the center-of-mass energy squared. In general the value of $\eta$ is however arbitrary and must be constrained through the evaluation of higher order corrections to impact factors. It goes without saying that Eq.~\eqref{eq:23} gives nothing but reggeization of the gluon.

\section{Amplitude generation for the high energy effective action}
\label{sec:numerical-methods}

\begin{figure}[th]
  \centering
  \includegraphics[width=.8\textwidth]{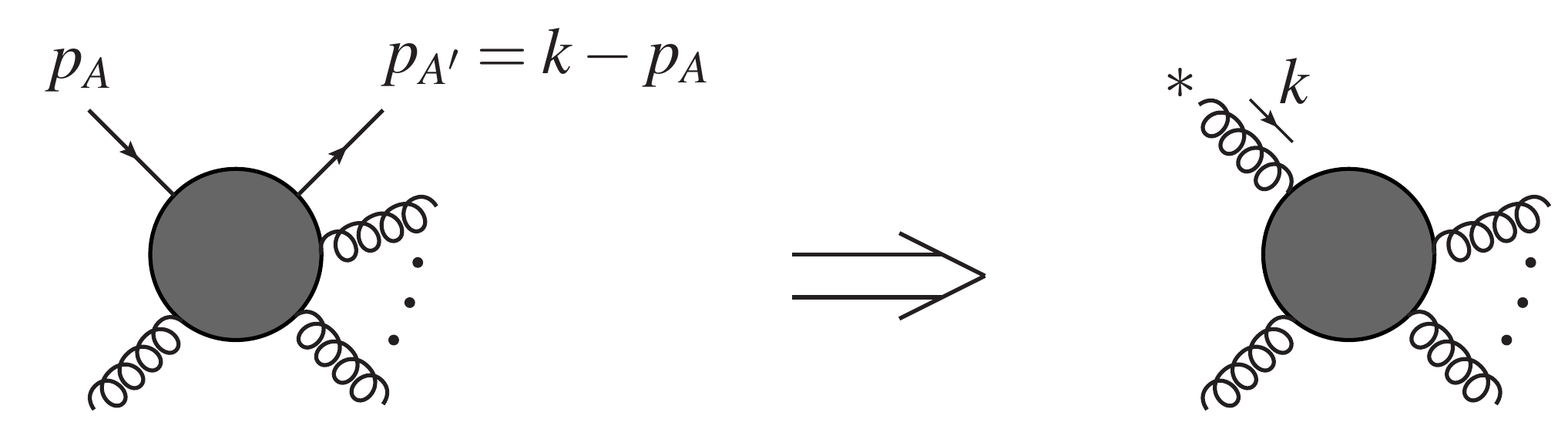}
  \caption{Left: A QCD scattering amplitude of arbitrary many external QCD states. A certain quark-antiquark pair is identified. Following the prescription given in the text, it is reduced to a reggeized gluon which then yields a corresponding scattering amplitude of a reggeized gluon with arbitrary many external partons [figure taken from \cite{vanHameren:2017hxx}]}
  \label{fig:vHam}
\end{figure}
For successful phenomenology of the QCD high energy limit, an
efficient determination of high energy factorized QCD scattering
amplitudes is of high importance. An important step in this direction
has been achieved through the parton-level event generator
\textsc{KaTie} \cite{vanHameren:2016kkz}. In a nut-shell one uses the
fact that conventional collinear tree-level scattering amplitudes can
be very effectively determined via numerical Dyson-Schwinger
recursion.  The state of a reggeized gluon is then generated through
an auxiliary quark-antiquark pair, which allows to obtain the
reggeized gluon state through a conventional colllinear scattering
amplitude. In more detail: Due to the kinematic constraint
Eq.~\eqref{eq:3}, the momenta of reggeized gluons satisfy
$k \cdot n^\pm = 0$. Relating light-cone directions of reggeized
gluons to actual momenta in a scattering process we have
$n^\pm = \sqrt{2} \frac{p_{1,2}}{\sqrt{p_1 \cdot p_2}}$ and a
reggeized gluon momentum can be parametrized as
\begin{align}
  \label{eq:24}
  k^\mu & = x p_1^\mu + {\bf k}^\mu, & k \cdot n^+ & = 0.
\end{align}
To generate the scattering amplitude of an off-shell reggeized gluon
with quarks and gluons, the reggeized gluon is then embedded into a
purely partonic scattering amplitude with the help of an auxiliary
quark-antiquark pair with momenta $p_A^\mu$ and $p_{A'}^\mu$. Both
auxiliary momenta  are  light-like  $p_A^2 = 0 =p_{A'}^2$. In more
detail one defines
\begin{align}
  \label{eq:25}
  p_A^\mu & = \Lambda p_1^\mu + \alpha p_2^\mu + \beta {\bf k}.
\end{align}
The requirement $p_A^2 = 0$ fixes then $\alpha =  \frac{\beta^2 {\bf k}^2}{\Lambda (p + q)^2}$  while  $p_{A'}^2 = 0 = (p_A - k)^2$ yields $\beta = \frac{1}{1 + \sqrt{1 - x/\Lambda}}$. In this way the momenta $p_A$ and $p_{A'}$ fulfill the important requirement of light-likeness for all values of $\Lambda$ as well as $p_A + p_{A'} = k$. The desired scattering amplitude of a single reggeized gluons with arbitrary many partons is then obtained from the limit\footnote{We use the symbol $r^*$ to denote the external state of an (off-shell) reggeized gluon}
\begin{align}
  \label{eq:26}
\mathcal{M}\left (0 \to r^*(xp_1 + {\bf k}) + X \right) & = \lim_{\Lambda \to \infty}  \frac{|{\bf k}|}{\Lambda} \cdot \mathcal{M} \left(0 \to q(p_A(\Lambda)), \bar{q}(p_{A'}(\Lambda)) + X\right).
\end{align}
Within the resulting scattering amplitude, it is then straightforward to distinguish gluons which are emitted from the auxiliary quark line from those which are not. The latter play the r$\hat{\text{o}}$le of the gluons which are generated by the QCD Lagrangian within the Lagrangian of the high energy effective action Eq.~\eqref{eq:2}. The former yield the contributions stemming from the induced Lagrangian. To make this relation  more explicit, it is useful to consider a typical propagator of the auxiliary quark line. With a generic momentum in the auxiliary quark line given by
\begin{align}
  \label{eq:27}
  p^\mu & = (\Lambda + x_p) p_1^\mu + y_p p_2^\mu + {\bf p},
\end{align}
 one then defines
\begin{align}
  \label{eq:29}
  p' & = p - p_{A'} = p - k +  p_A \notag \\
   & = (2 \Lambda - x) p_1 + (y_q + \alpha)p_2 + {\bf p} - {\bf k} (1-\beta),
\end{align}
as the momentum flowing in the (auxiliary) quark propagator, if the external quark $p_A$ would given the momentum $k$ and the anti-quark momentum $0$, {\it i.e.} the total momentum of the gluons emitted by the quark sum up to the reggeized gluon momentum.  $p'$ is therefore the total momentum  of gluons``so far'' emitted from auxiliary quark line. In particular
\begin{align}
  \label{eq:30}
\lim_{\Lambda \to \infty}  p_1 \cdot p' & = y_p 2 p_1 \cdot q.
\end{align}
 One then finds for the  quark propagator with momentum $p$ in the limit $\Lambda \to \infty$:
\begin{align}
  \label{eq:28}
 \lim_{\Lambda \to \infty}  \frac{{p\!\!\!\!/}}{p^2} & =  \lim_{\Lambda \to \infty}  \frac{(\Lambda + x_p) {p \!\!\!\!/}_1 + y_p {p \!\!\!\!/}_2 + {\bf p \!\!\!\!/}}{ 2(\Lambda + x_p) y_p p_1 \cdot p_2 + {\bf p}^2} =  \frac{{p \!\!\!\!/}_1}{2 p_1\cdot p'} = \frac{{n \!\!\!\!/}^+}{2 p^{+'}} \, .
\end{align}
Making now use of well-known identities such as
\begin{align}
  \label{eq:31}
 {n \!\!\!\!/}^+  \gamma_\mu  {n \!\!\!\!/}^+ & = 2 n^\mu \cdot {n \!\!\!\!/}^+,
\end{align}
one  re-obtains from these contributions the induced
vertices\footnote{In more detail: one first obtains the term
  corresponding to one particular ordering of the operator
  Eq.~\eqref{eq2:efflagrangian}; summing over all possible
  orderings,  the full color structure of
  Fig.~\ref{fig:3} is obtained} of the high energy effective action,
Fig.~\ref{fig:3}.  The whole procedure can be generalized in a
straightforward manner to the case of two reggeized gluons with
opposite {\it i.e.} plus- and minus-polarization. In this way it is
then possible to generate all tree-level amplitudes of the high-energy
effective action. This is of particular use, since there exists very
powerful algorithms which allow to generate the underlying scattering
amplitudes numerically. It is therefore correct to claim that the
problem of efficiently computing gauge invariant tree level scattering
amplitudes within high energy factorization in the region of phase
space where one reggeized gluon exchange provides a suitable
approximation, has been completely solved in recent years, both
analytically and numerically. For details we refer the interested
reader to the original papers \cite{vanHameren:2012uj,
  vanHameren:2013csa, vanHameren:2012if, vanHameren:2017hxx,
  vanHameren:2014iua,Bury:2015dla,vanHameren:2015bba,vanHameren:2016bfc}. The
method has been used in the recent past for a large number of
phenomenological studies, but also for theoretical exploration, see
for instance the calculation of transverse momentum dependent
splitting kernels in \cite{Hentschinski:2017ayz}.

\section{High energy effective action and high parton densities}
\label{sec:high-energy-effect-4}

An area of high phenomenological interest is given by the physics of
high parton densities: occurrence of high parton densities is closely
related to the physics of gluon saturation which is assumed to the be
relevant mechanism to ensure unitarity of QCD scattering amplitudes in
the high energy limit. The bulk of theoretical efforts in recent years
to describe this region of phase space within high energy
factorization, can be summarized under the term ``Color Glass
Condensate effective field theory'' (see \cite{Gelis:2010nm} for a review). In the following we summarize the essential results of \cite{Hentschinski:2018rrf} which demonstrate of these efforts with the high energy effective action. 

\subsection{A special parametrization of the gluonic field}

The starting point for this exploration has already been laid in the
original publication \cite{Lipatov:1995pn,Lipatov:1996ts} by Lipatov himself: there it has
been noted that while a shift in the gauge field
$v^\mu \to v^\mu + n_-^\mu A_+/2 + n_+^\mu A_-/2$ provides a
possibility to avoid the appearance of a direct transition vertex
between reggeized gluon field and gluon field, Fig.~\ref{fig:3}.a,  it mixes two fields with different
gauge transformation properties: $v_\mu$ transforms as a gauge field, while $A_\pm$ is by definition
invariant under local gauge transformations. The solution put forward in \cite{Lipatov:1995pn,Lipatov:1996ts} suggests the following parametrization of the gluonic field:
\begin{align}
  \label{eq:para0}
  V^\mu(x) & =
 v^\mu(x) + \frac{n_+^\mu}{2}   B_-(x) + \frac{n_-^\mu}{2}   B_+(x)  \, ,
\end{align}
where 
\begin{align}
  \label{eq:1B}
  B_\pm[v_\mp] = U[v_\mp] A_\pm U^{-1}[v_\mp] \,.
\end{align}
and (inverse) Wilson line operators are   defined as 
\begin{align}
  \label{eq:U}
  U[v_\pm] &=  \frac{1}{1 + \frac{g}{\partial_\pm} v_\pm}, &  
U^{-1}[v_\pm]
  & = 1 + \frac{g}{\partial_\pm} v_\pm \, .
\end{align}
For the above composite field $B_\pm[v_\mp]$, one finds the following
gauge transformation properties:
\begin{align}
  \label{eq:deltaterm}
  \delta_L B_\pm & = \delta_L U[v_\mp] A_\pm U^{-1}[v_\mp] + U[v_\mp] A_\pm\delta_L U^{-1}[v_\mp]  = \left[g B_\pm, \chi_L \right]\,,
\end{align}
and as a consequence  the shifted gluonic field
Eq.~\eqref{eq:para0} transforms like a gluonic field:
\begin{align}
  \label{eq:deltaterm2}
    \delta  V_\pm  & = \left[D_\pm, \chi \right] + [g B_\pm, \chi] =  \left[D_\pm + g B_\pm, \chi \right].
\end{align}
 In the following we will
use the above parametrization of the gluonic field to expand the high
energy effective action for the quasi-elastic case around the
reggeized gluon field $A_+$ which we treat as a strong classical background
field $g A_+ \sim 1$. We therefore consider the high energy effective action for the quasi-elastic case
\begin{align}
  \label{eq:2qe}
  S_{\text{eff}}^{\text{q.e.}} & = S_{\text{QCD}} + S_{\text{ind.}}^{\text{q.e.}}
\end{align}
with
\begin{align}
  \label{eq:4qe}
  S_{\text{ind.}}^{\text{q.e.}} & = \int d^4 x\, \text{tr}  \left( \left\{ T_-[v] - A_- (x) \right\}  \partial^2 A_+(x)\right]\,.
\end{align}
Keeping fields $A_+$ to all orders and expanding in quantum fluctuations $v_\mu$ and $\psi$, $\bar{\psi}$  to quadratic order, we obtain
\begin{align}
  \label{eq:5qe}
   S_{\text{eff}}^{\text{q.e.}} & = \int d^4x \left[\mathcal{L}_0 + \mathcal{L}_1 - \text{tr}\left( A_-\partial^2 A_+  \right)\right] + \mathcal{O}(v_\mu^3),
\end{align}
with the kinetic term of the gluonic  and quark field
\begin{align}
  \label{eq:l0}
  \mathcal{L}_0 & =  \text{tr} \left( -v^\mu [g_{\mu\nu} \partial^2 - \partial_\mu \partial_\nu] v^\nu\right) + \bar{\psi} i {\partial \!\!\!\!/}  \psi,
\end{align}
and the quadratic terms which describe interaction with the reggeized gluon field,
\begin{align}
  \label{eq:Lagrangian1X}
 \mathcal{L}_1 & = g\cdot  \bigg\{ \frac{i}{2}\bar{\psi} {n \!\!\!\!/}_- A_+ \psi +  \text{tr} \bigg[
 \partial_- v_\mu [ A_+, v^\mu] 
+  2 \partial_\mu v_- [v^\mu, A_+]   +
\notag \\
& \hspace{2cm}
  +    
  \partial^2 v_-   \left[\left(\frac{1}{\partial_-}v_- \right),  A_+ \right] -  v_- \left(\frac{1}{\partial_-}v_- \right) \partial^2 A_+ \bigg] \bigg\}\,.
\end{align}

\subsection{Resummation of a dense reggeized gluon field}
\label{sec:dense}

The quasi-elastic effective action allows then for the extraction of quark-quark-reggeized gluon (QQR) and gluon-gluon-reggeized gluon vertices, 
\begin{align}
  \label{eq:QQR}
  \parbox{2.8cm}{\includegraphics[width=2.8cm]{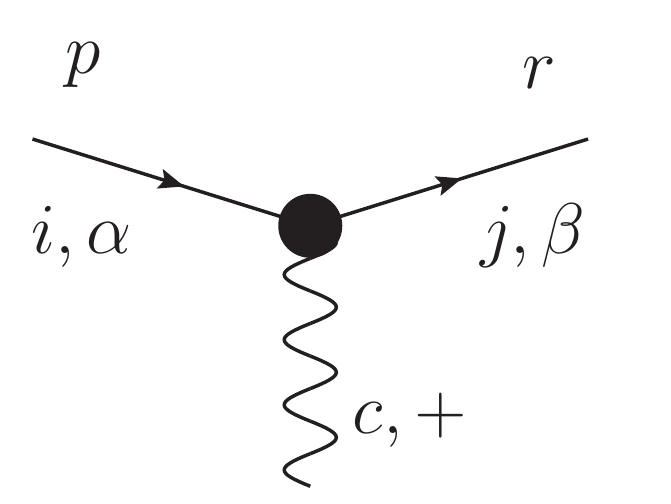}} & = -ig t^c_{ji} \Gamma_{\beta\alpha}(r,p) \int d^4z \, e^{-iz\cdot(p-r)} A^c_+(z),  \notag \\
 &   \Gamma_{\beta\alpha}(r,p) = -\frac{1}{2} {n \!\!\!\!/}^+_{\alpha\beta}\,,
\\
  \label{eq:GG}
  \parbox{2.8cm}{\includegraphics[width=2.8cm]{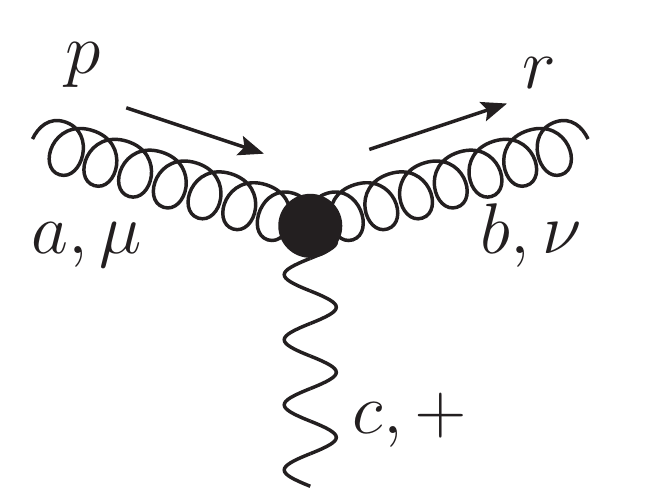}} & = -ig T^c_{ba} \Gamma^{\nu\mu}(r,p)  \int d^4z \, e^{-iz\cdot(p-r)} A^c_+(z),  \notag \\
     \Gamma_+^{\nu\mu}(r, p)  &= p^+ g^{\mu\nu} - (n^+)^\mu p^\nu - (n^+)^\nu r^\mu  + \frac{r \cdot p}{p^+}  (n^+)^\mu (n^+)^\nu \,,
\end{align}
with $T^c_{ab} = -if^{abc}$. The QQR and GGR vertices obeys various properties:
\begin{align}
  \label{eq:9qe}
  \Gamma_{\beta\gamma'}(r,p) {n \!\!\!\!/}_{\gamma'\gamma} &= 0   =   {n \!\!\!\!/}_{\beta\beta'}  \Gamma_{\beta'\gamma}(r,p), 
\notag \\
 \Gamma_{\beta\gamma}(r,k) {k \!\!\!\!/}_{\gamma\gamma'} \Gamma_{\gamma'\alpha}(k,p)
 & = 
-p^+ \Gamma_{\beta\alpha}(r,p) ,
 \end{align}
 and
\begin{align}
  \label{eq:9ge}
  n^+_\nu \cdot \Gamma^{\nu\mu}_+(r,p) & = 0 =    \Gamma^{\nu\mu}_+(r, p) \cdot n^+_\mu , 
\notag \\
  \Gamma^{\nu\alpha}_+(r,k)\cdot  (-g_{\alpha\alpha'} )\cdot  \Gamma^{\alpha'\mu}_+(k,p) & =   -p^+   \Gamma^{\nu\mu}_+(r,p)\, ,
       \notag \\
r_\nu  \cdot  \Gamma^{\nu\mu}_+(r, p)  & = 0 =    \Gamma^{\nu\mu}_+(r, p) \cdot p_\mu. 
\end{align}
Note that the above GGR-vertex was already obtained in \cite{Lipatov:1995pn}.  A
potential disadvantage of this vertex, pointed out in \cite{Lipatov:1995pn},
is that it can yield individual Feynman diagrams which contain
singularities in overlapping channels and therefore violate the
Steinmann-relations \cite{Steinmann}. Since the vertex is obtained
from an effective action which explicitly respects the Steinmann
relations, such effects must cancel at the level of the
observable. The properties Eq.~\eqref{eq:9qe} are of great use for the
resummation of a strong reggeized gluon field $g A_+ \sim 1$ to all
orders. Such an exercise has been carried out in
\cite{Hentschinski:2018rrf}. As a result one obtains the following effective
vertices which sum up the interaction of a quark and a gluon with an
arbitrary number of reggeized gluon fields:
\begin{align}
\parbox{2.8cm}{\includegraphics[width=2.7cm]{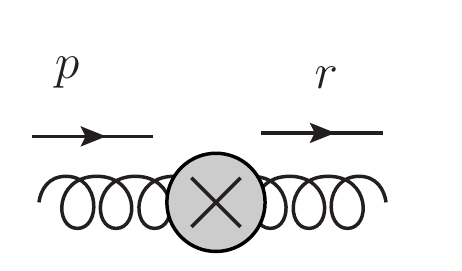}} & = 
\tau_{G,\nu\mu}^{ab}(p, -r) =  
  - 4 \pi \delta(p^+ - r^+)  \Gamma_{\nu\mu}(r,p)  e^{-i x_0^+(p^- - r^-)}
 \notag \\  & \hspace{-2cm}
 \cdot \int d^2 {\bf z} e^{i {\bf z} \cdot ({\bf p} - {\bf r})}   
\bigg[ \theta(p^+)  \left[ U^{ba}({\bf z}) - \delta^{ab} \right]-
              \theta(-p^+) \left[  [U^{ba}({\bf z})]^\dagger  -
              \delta^{ab} \right]\bigg],
\label{eq:finally}
\\
 \parbox{2.8cm}{\includegraphics[width=2.7cm]{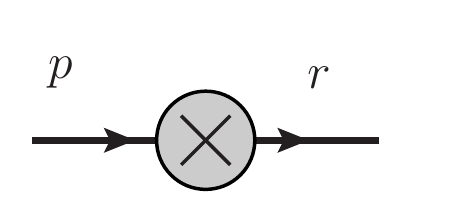}} & = \tau_{F}(q,-r)=
  2 \pi \delta(p^+ - r^+) {n \!\!\!\!/}^+  e^{-i x_0^+(p^- - r^-)}  
\notag \\
& \hspace{-1cm}\cdot  \int d^2 {\bf z} e^{i {\bf z} \cdot ({\bf p} - {\bf r})}
\bigg[ \theta(p^+)  \left[ W({\bf z}) - 1 \right]-  
\theta(-p^+) \left[  [W({\bf z})]^\dagger  - 1 \right]\bigg]\,.
 \end{align}
To write down the above expressions,  we introduced Wilson lines in the adjoint
\begin{align}
  \label{eq:Uab}
   U^{ab}({\bf z}) & =    \mathrm{P} \exp \left(-\frac{g}{2}\int_{-\infty}^\infty dz^+ \tilde{A}_+ \right), &   \tilde{A}_+ & = -iT^c_{ab} A^c_+\,,
\end{align}
and the fundamental representation
\begin{align}
  \label{eq:W}
  W({\bf z}) & =   \mathrm{P} \exp \left(-\frac{g}{2}\int_{-\infty}^\infty dz^+ {A}_+ \right), &   {A}_+ & = -it^c_{ij} A^c_+\,.
\end{align}
The above vertices Eqs.~\eqref{eq:finally} allow to
construct partonic propagators in the presence of a (reggeized gluon)
background field and can be compared with corresponding propagators
derived for the resummation of a background field in the light-cone
gauge \cite{McLerran:1994vd,Baltz:2001dp,Gelis:2001da,Balitsky:2001mr, Ayala:2016lhd,Ayala:2017rmh}. It turns that the
regarding propagators agree with each other, if the gluonic field is taken in
the same gauge, {\it i.e.} light-cone gauge. The interpretation is
however slightly different for the Wilson line which resum
gluonic background field in light-cone gauge and reggeized gluon field
respectively. For a more detailed discussion we refer the interested
reader to \cite{Hentschinski:2018rrf}. As a first application it was
possible to use the vertices
Eq.~\eqref{eq:finally} to re-derive 
the leading order Balitsky-JIMWLK equation from the high energy
effective action: Calculating quantum fluctuations of an ensemble of
Wilson lines of the reggeized gluon fields
\begin{align}
  \label{eq:44}
  U^{ab}({\bf z}) & = 2 \text{tr} \left[ t^a W({\bf z})t^b  W^\dagger({\bf z})\right]\,,
\end{align}
one finds for the evolution with respect to the regulator used for the high energy singularities, the Balitsky-JIMWLK equation 
\begin{align}
  \label{eq:2.6}
  - \Lambda_a \frac{d}{d \Lambda_a} \left[W({\bf x}_1) \otimes \ldots \otimes W({\bf x}_n) \right]
&=
\sum_{i,j = 1} H_{ij}  \left[W({\bf x}_1) \otimes \ldots \otimes W({\bf x}_n) \right]\,,
\end{align}
with the Balitsky-JIMWLK Hamiltonian
\begin{align}
  \label{eq:2.7}
  H_{ij} & = \frac{\alpha_s \Gamma^2(1+\epsilon)}{2 \pi^2 \Gamma(1-\epsilon)} \left(\frac{4 }{\pi \mu^2} \right)^\epsilon   \int d^{2+2\epsilon} {\bf z} 
\frac{({\bf x}_i - {\bf z}) \cdot ({\bf x}_j - {\bf z})}
{[({\bf x}_i - {\bf z})^2]^{1+\epsilon} [({\bf x}_j - {\bf z})^2]^{1+\epsilon}}
\notag \\
& \hspace{2.5cm}
\left[ 
T^a_{i, L} T^a_{j, L} + T^a_{i, R} T^a_{j, R} - U^{ab}({\bf z}) 
\left(T^a_{i, L} T^b_{j, R} + T^a_{j, L} T^b_{i, R}    \right)
\right], 
\end{align}
where $T^a_{L,i}$ and $T^a_{R,j}$ are the group generators acting
to the left (L) or to the right (R) on the Wilson line $W({\bf x}_i)$,
\begin{align}
  \label{eq:2.5}
  T^a_{L,i} [W({\bf z}_i)]&  \equiv t^a W({\bf z}_i), &
 T^a_{R,i} [W({\bf z}_i)]&  \equiv W({\bf z}_i) t^a.
\end{align}

\section{Conclusions}
\label{sec:conclusions}

The topics covered in this short review provide apparently only a
small subset of results achieved within the high energy effective
action (with the choice heavily biased by the author's own
work). Among the results which we did not review here, are the
effective action for Regge processes in gravity \cite{Lipatov:2011ab},
earlier versions of an high energy effective action
\cite{Lipatov:1991nf,Kirschner:1994gd,Kirschner:1994xi} as well as the
determination of the $3 \to 3$ reggeized gluon transition kernel,
needed for the NLO Bartels-Kwiecinski-Praszalowicz equation in
\cite{Bartels:2012sw} and the NLO corrections to the Mueller-Tang jet
impact factor
\cite{Hentschinski:2014lma,Hentschinski:2014bra,Hentschinski:2014esa}. Another
line of research has been developed in a series of papers
\cite{Braun:2006sk,Braun:2009ki,Braun:2011qm,Braun:2011it,Braun:2012gs,Braun:2013xoa,Braun:2014rfa,Braun:2016xvx,Braun:2019gjy}
dedicated to the study of production processes at central rapidities
in the presence of multiple reggeized gluon exchange.  The relation of
the high energy effective action to the Color Glass Condensate
approach as well as the further development of the underlying
formulation has been explored in
\cite{Bondarenko:2017vfc,Bondarenko:2017ern} and continued in
\cite{Bondarenko:2018pvv,Bondarenko:2018geo,Bondarenko:2018kqs,Bondarenko:2018eid,Bondarenko:2019esj}. In
parallel to the high energy effective action for gluons, the
corresponding effective action for quarks has been explored in series
of papers, at first mainly focusing on phenomenological applications
in combination with transverse momentum dependent parton distribution
functions \cite{Saleev:2008cd,Saleev:2009et,Kniehl:2010iz,Kniehl:2011hc,Saleev:2012hi,Saleev:2012np,Nefedov:2012cq,Nefedov:2013hya,Saleev:2013yta,Nefedov:2013qya,Nefedov:2013ywa,Kniehl:2014qva,Nefedov:2015ara,Karpishkov:2014epa,Nefedov:2014qea,Karpishkov:2016hnx,Karpishkov:2017kph,Nefedov:2018vyt}, while recent efforts address the determination of
higher order perturbative corrections
\cite{Nefedov:2017qzc,Nefedov:2019mrg}.
\\

When I met Lev Lipatov in 2006, I had just started working on my
doctoral thesis at the II. Institute for Theoretical Physics at
Hamburg University. While working with him was at first a challenge, I soon realized what a great teacher he was and how many
insights he was generously sharing with me.  I must admit that I could
not always follow all of his explanations and that sometimes it took
me years until I was able to finally understand the full impact of one
or the other comment he made to me during these days. While the high energy
effective action might be not among the achievements Lipatov is most famous
for, I am deeply convinced that it constitutes an important part of
his legacy. It allows us to benefit from his deep insights into the Regge limit of QCD even after he went away from us.

\section*{Acknowledgements}

Support by Consejo Nacional de Ciencia y Tecnolog{\'\i}a grant number A1 S-43940 (CONACYT-SEP Ciencias B{\'a}sicas) is gratefully acknowledged.

\bibliographystyle{unsrt}    
\bibliography{paper}

\end{document}